\documentclass[11pt]{article}

\usepackage[bookmarks=false,hidelinks]{hyperref}
\usepackage{xcolor}
\usepackage{cite}

\usepackage{caption}
\usepackage{amsmath,amssymb}
\usepackage{mathpazo}
\usepackage{tocloft}
\usepackage{siunitx}

\usepackage{grffile} 

\usepackage{empheq}

\usepackage{tcolorbox}

\renewcommand{\d}{{\rm d}} 

\usepackage{color}
\definecolor{blue}{rgb}{0.00,0.00,0.95}

\begin{document}

\centerline{\bf{\LARGE{Learning dynamical models of single}}} 
\centerline{\bf{\LARGE{and collective cell migration: a review}}}
\vspace{0.4 cm}

\centerline{{David B. Br\"uckner$^{1,*}$ and Chase P. Broedersz$^{2,3,\dagger}$}}
$^1$Institute of Science and Technology Austria, Am Campus 1, 3400 Klosterneuburg, Austria \\
$^2$Department of Physics and Astronomy, Vrije Universiteit Amsterdam, 1081 HV Amsterdam, The Netherlands \\
$^3$Arnold Sommerfeld Center for Theoretical Physics and Center for NanoScience, Department of Physics, Ludwig-Maximilian-University Munich, Theresienstr. 37, D-80333 Munich, Germany \\
$^*$david.brueckner@ist.ac.at, 
$^\dagger$c.p.broedersz@vu.nl


Single and collective cell migration are fundamental processes critical for physiological phenomena ranging from embryonic development and immune response to wound healing and cancer metastasis. To understand cell migration from a physical perspective, a broad variety of models for the underlying physical mechanisms that govern cell motility have been developed. A key challenge in the development of such models is how to connect them to experimental observations, which often exhibit complex stochastic behaviours. In this review, we discuss recent advances in data-driven theoretical approaches that directly connect with experimental data to infer dynamical models of stochastic cell migration. Leveraging advances in nanofabrication, image analysis, and tracking technology, experimental studies now provide unprecedented large datasets on cellular dynamics. In parallel, theoretical efforts have been directed towards integrating such datasets into physical models from the single cell to the tissue scale with the aim of conceptualizing the emergent behaviour of cells. We first review how this inference problem has been addressed in both freely migrating and confined cells. Next, we discuss why these dynamics typically take the form of underdamped stochastic equations of motion, and how such equations can be inferred from data. We then review applications of data-driven inference and machine learning approaches to heterogeneity in cell behaviour, subcellular degrees of freedom, and to the collective dynamics of multicellular systems. Across these applications, we emphasize how data-driven methods can be integrated with physical active matter models of migrating cells, and help reveal how underlying molecular mechanisms control cell behaviour. Together, these data-driven approaches are a promising avenue for building physical models of cell migration directly from experimental data, and for providing conceptual links between different length-scales of description.


\tableofcontents

\section{Introduction}
\label{sec:intro}

The vast majority of cells in our body do not move around -- but when they do, it is for an important reason: migrating cells shape you, they can protect you, but also harm or even kill you. In development, cells actively migrate to be at the right place at the right time, shaping the early embryo~\cite{Franz2002,Scarpa2016}. Later on, while most cells become sedentary, immune cells have a remarkable ability to migrate through the tightest pores to hunt down pathogens, protecting you from diseases~\cite{Luster2005}. Furthermore, all cells retain the ability to switch to a migratory mode, allowing them to efficiently close wounds~\cite{Li2011a,Bainbridge2013}. However, this ability is hijacked by cancer cells, which migrate during metastasis to spread to other organs~\cite{Friedl2003,Yamaguchi2005,Paul2017}. 

The underlying processes required to make a cell move are determined by a broad variety of physical phenomena: the polymer physics of cytoskeletal filaments~\cite{Bausch2006,Kasza2007,Fletcher2010,Broedersz2014}, the reaction-diffusion dynamics of signaling molecules~\cite{Maiuri2015,CallanJones2016}, and the active mechanics of acto-myosin contraction~\cite{Kruse2005,Prost2015,Julicher2018}. The cellular motility machinery integrates these physical processes to push forward the cell membrane, giving rise to overall motion of the cell. Much of this machinery is highly conserved across organisms and tissues~\cite{Pollard2003}, giving hope that understanding the physics of these processes will lead to a general understanding of cell motility. However, while much progress has been made in understanding each of these biophysical aspects, how these integrate to generate behaviours at the scale of the cell as a whole remains the subject of current research. 

An exciting perspective is therefore whether physics can go beyond explaining the physical components of cellular systems and provide conceptual and predictive frameworks to describe the emergent behaviour of cells as a whole. To accomplish this, we need to connect physical modelling approaches across scales and understand how they interplay at the system level. To achieve such connections in systems with such daunting inherent complexity, data-driven theoretical approaches that connect directly to experimental data are emerging as a fruitful and promising avenue. Put simply, such data-driven approaches aim to solve the inverse problem of determining an effective physical description of a system from data. Indeed, in recent years, a number of studies have started developing data-driven approaches to learn dynamical models of stochastic cell migration directly from experimental data. This includes a wide variety of inference approaches using stochastic inference, machine learning and dimensional reduction to infer how cells interact with their environment and with each other. This field is currently at a unique crossroad: due to advances in nanofabrication, image analysis and tracking technology, experimental studies now yield unprecedented large data sets on cellular phenotypes; and at the same time, there is an increasing pivot among theoreticians to interact directly with experimental data and apply tools such as machine learning and physics-guided inference approaches to learn from data.

\begin{figure}[h!]
	\includegraphics[width=0.8\textwidth]{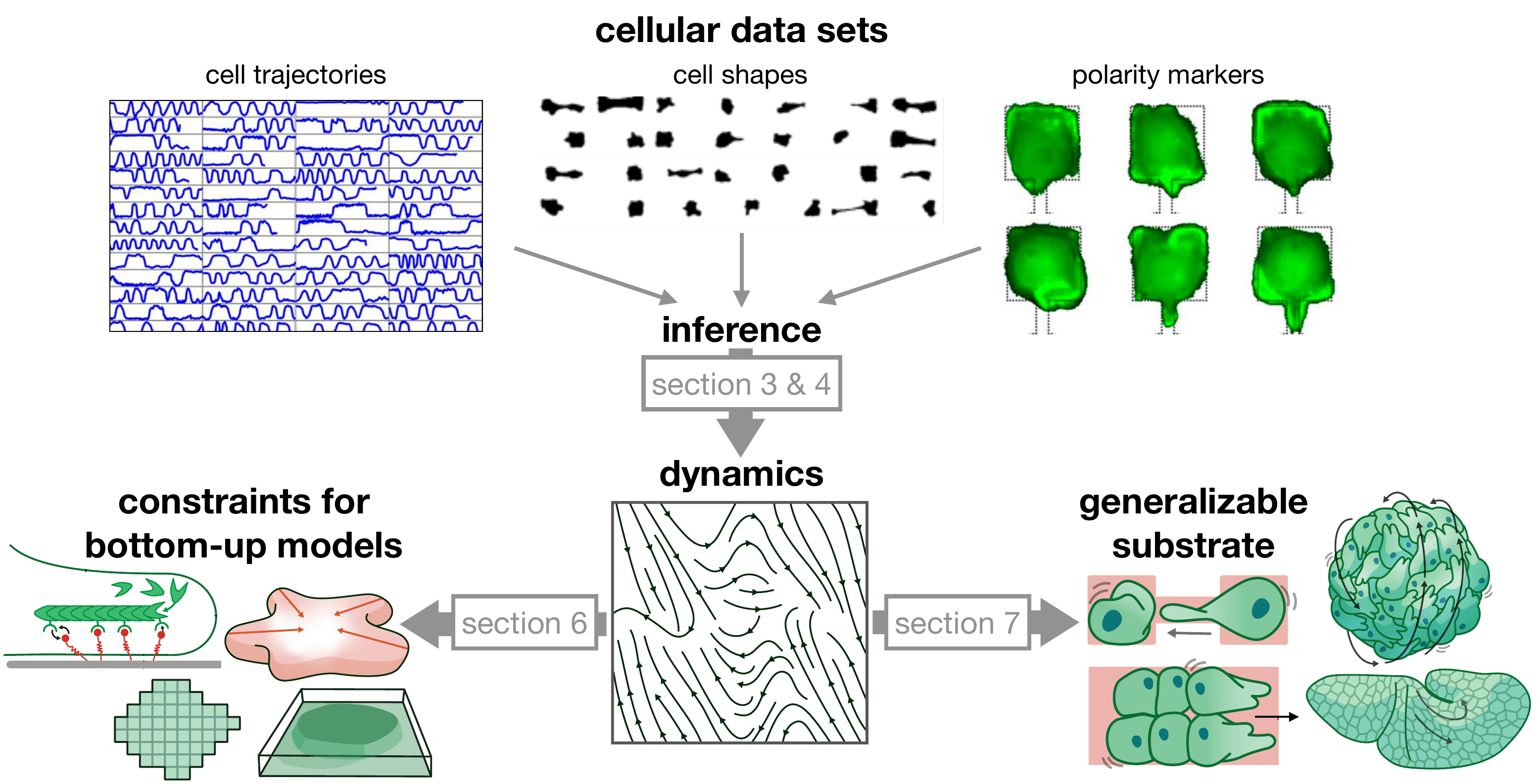}
	\centering
		\caption{
		\textbf{Conceptual approach of learning data-driven models from cellular data sets.} Cellular data sets such as cell trajectories, cell shapes or intracellular markers serve as input to model inference (here shown for the example of a confined cell~\cite{Brueckner2019,Fink2019}). This provides a dynamical systems representation of behaviour, providing constraints for bottom-up models, and a generalizable basis for more complex systems.
				 }
	\label{fig:overview}
\end{figure}

In this article, we take stock of these recent advances and the outstanding challenges in learning dynamical models of the stochastic behaviour of single and collective cell migration directly from experimental data (Fig.~\ref{fig:overview}). A key challenge for these approaches is to connect to more classical biophysics models of cell migration, including soft matter, hydrodynamic, and mechanical theories. These include mechanistic models at the single cell level (active gel theory, phase field models, Cellular Potts Models); and active matter models at the collective scale (active hydrodynamics, active particle models, vertex models). We discuss how inference from data can be connected and integrated with these physical approaches, and how it may provide bridges to connect diverse modelling approaches into a coherent framework for cell migration. As the diversity, accuracy, dimensionality, and size of these cellular datasets is rapidly increasing, we expect such data-driven approaches to play an increasingly important role in building physical models of single and collective cell behaviour.

First, we provide a perspective of how we envision learning cell dynamics at the behavioural level will advance our understanding of cell migration (section~\ref{sec:behaviour}). We then review data-driven approaches to learn the dynamics of single migrating cells (section~\ref{sec:single_cell}), and summarize the technical aspects of performing stochastic inference from cell trajectories (section~\ref{sec:inference}). Furthermore, we will review how these approaches have been extended to give insight into the variability of cell behaviours in time and across individuals (section~\ref{sec:variability}). In section~\ref{sec:mechanism}, we provide a perspective on how data-driven approaches to emergent cell dynamics can be connected to underlying molecular mechanisms. Zooming out from the single cell level, we then review data-driven approaches to describe the interactions between cells (section~\ref{sec:collective}). These examples demonstrate how combining advances in physical modeling, inference methods, and high-throughput experimental approaches can help reveal the underlying physics of what makes a cell move.

\section{Cell migration at the behavioural level}
\label{sec:behaviour}

In multicellular organisms, individual cells migrate to execute functional tasks. Thus, cells are programmed to perform certain behaviours, including for example net motion (migration), changes in cell shape (morphodynamics), exerting forces on the extra-cellular environment (traction forces), adaptation to external signals (stimulus response), or the degradation of surrounding matrix polymers (proteolysis). What all these examples have in common, is that they are performed at the scale of the whole cell and often take place on long time-scales. Here, we refer to `long time-scales' as those time intervals which are long compared to typical time-scales of intra-cellular processes, such as the polymerization of an actin filament or the life-time of a focal adhesion. On these large time- and length-scales, cellular behaviour emerges as a consequence of a large number of intra-cellular processes operating simultaneously. From the point of view of physical models of cell migration, this complexity means that molecularly reductionistic approaches are very challenging: precise knowledge of one or several particular signalling processes and all the associated parameters may not be predictive for the whole-cell behaviour. This is because whole-cell behaviours integrate many processes, and developing quantitative models for all of these at ones is unfeasible.

To circumvent this problem, minimal physical models are often employed, which seek to identify the key mechanisms at play and integrate them into a physical model. Such approaches typically postulate specific mechanisms and explore their implications. We refer to these approaches as \textbf{bottom-up models} in this review, since these approaches start by postulating a set of rules to describe the various components of a cell and then seek to predict the emerging behaviour. Examples for this are Cellular Potts models~\cite{Graner1992,Glazier1993,Segerer2015,Albert2014,Albert2016b,Albert2016,Albert2016a,Thuroff2019a,Goychuk2018}, phase field models~\cite{Kockelkoren2003,Ziebert2011,Shao2012a,Marth2014,Camley2014a,Bertrand2020}, as well as the molecular clutch model~\cite{Chan2008,Elosegui-Artola2018}, active gel theories~\cite{Kruse2006,Callan-Jones2008,Hawkins2009,Hawkins2011,Blanch-Mercader2013,Khoromskaia2015,Bergert2015,Recho2019,Lavi2020} and models coupling actin flow, polarity cues, and focal adhesion dynamics~\cite{Gracheva2004,Doubrovinski2011,Vanderlei2011,Maiuri2015,CallanJones2016,Camley2017b,Ron2020a,Sens2020,Hennig2020,Schreiber2021}. We will review these approaches in more detail in section \ref{sec:mechanism}. However, applying these types of models directly to experimental observations is challenging: depending on the implementation, these models may have many parameters that are difficult to constrain based on experimental data. To avoid this, models are frequently tailored to capture a particular aspect of the data, but it has often remained difficult to capture the full long time-scale dynamics of the cells, or how these dynamics adapt to external inputs.

An alternative to mechanistic models are data-driven \textbf{top-down approaches}, which systematically constrain model candidates using experimental data. Naturally, top-down approaches tend to provide a more phenomenological description of the system since they are typically based on experimental data at the cellular or tissue scale rather than the molecular scale. An example of a top-down approach are models inferred directly from measured cell migration trajectories. The resulting phenomenological description based on such data therefore effectively coarse-grains over the molecular detail. Generally speaking, phenomenological theories in physics have often generated conceptual understanding that remained elusive in the reductionist approach, an idea that was famously articulated by Phil Anderson in his essay `More is different'~\cite{Anderson1972}. Indeed, different levels of description can be relevant at different time- and length-scales, suggesting that the molecularly reductionistic approach is not the only way of modelling a system, but phenomenological descriptions could be very helpful at the large time- and length-scales of cellular behaviours. Following this philosophy, we argue that top-down approaches form a promising direction to develop quantitative frameworks for cell behaviour. These approaches are generally characterized by the following properties: 
\begin{itemize} 
\item \textbf{Data-driven}: To provide a phenomenological description of a cellular system without reference to specific molecular processes, top-down approaches need to be constrained by experimental data. This can be achieved by employing data-driven inference techniques, which constrain postulated models using input data, meaning that high-quality quantitative datasets of cell behaviours are required.
\item \textbf{Unbiased}: A central idea in top-down approaches is that they should be agnostic with respect to the underlying molecular or mechanistic basis of the behaviour. Specifically, first a general model class is proposed based on symmetry or causality arguments. These models should be constrained by experimental data in a principled manner, rather than using pre-conceived intuition. The hope is that an unbiased approach can yield a more general description of cell behaviours, which could then be used to systematically constrain mechanistic, bottom-up frameworks. 
\item \textbf{Predictive}: While a given model may be constrained using data, it should then also be able to predict new observations beyond the data that were used to constrain it. Tests of predictive power have two distinct roles: firstly, making predictions for the same experimental data set used to constrain the model, but for statistics that were not explicitly used in the inference, allows testing whether the model provides a meaningful representation of the cellular behaviour. Secondly, performing predictions for new experiments tests the usefulness of the model to provide a generalizable basis for new systems. 
\end{itemize} 

There has been a recent surge of activitity in developing such data-driven, unbiased approaches in a number of other biological systems across scales, including protein folding~\cite{Pearce2019,Jumper2021}, chromosome organization~\cite{Imakaev2015,Abbas2019,Messelink2021} and dynamics~\cite{Gabriele2021,Mach2022,Brueckner2023}, neural systems~\cite{Schneidman2006,Tkacik2013,Genkin2020,Mlynarski2021} and animal behaviour~\cite{Berman2014,Brown2018,Stephens2008,Stephens2011,Costa2019}. In recent years, the abundance of data-sets of dynamical systems across the disciplines has been rapidly increasing. This has led to the development of data-driven methods to infer the underlying dynamics of complex systems directly from experimental data, including deterministic~\cite{Crutchfield1987,Daniels2015,Brunton2015,Champion2019,Chen2021} and stochastic trajectories~\cite{Siegert1998a,Ragwitz2001,Beheiry2015,PerezGarcia2018,Frishman2018,Boninsegna2018,Costa2019,Brueckner2020a,Ferretti2020,Dai2020a,Callaham2021,Huang2022}, spatially extended fields~\cite{Rudy2016,Borzou2021}, and morphological dynamics~\cite{Brown2018,Stephens2008,Stephens2011}. We argue that there are four key challenges that inference approaches for cell behaviour could help address, based on which we organize the structure of this review (Fig.~\ref{fig:overview}):
\begin{enumerate} 
\item Owing to the intrinsic stochasticity and variability of cell behaviours, a key challenge is to identify what constitutes a `typical' behaviour. Data-driven approaches could provide \textbf{analysis tools} for unbiased, quantitative characterization, classification, and observation of cellular behaviours. Examples for cellular readouts are cell persistence in 2D migration~\cite{Selmeczi2005,Maiuri2015}, transition times and occupancy probabilities in confined migration~\cite{Brueckner2019,Fink2019}, movement biases in directional migration~\cite{LoVecchio2020}, and collision outcomes in cell-cell interactions~\cite{Scarpa2013,Milano2016,Brueckner2021}. We discuss how data-driven approaches can provide readouts of typical behaviours (section~\ref{sec:single_cell}) and of their variability (section~\ref{sec:variability}).
\item Due to the emergent nature and underlying complexity of cell behaviours, it is often unclear what the right quantitative concepts are to describe a particular observed behaviour. Data-driven approaches could yield \textbf{conceptual frameworks} to think about cell behaviours by identifying underlying quantitative concepts that can be used to describe cell dynamics. Examples for such concepts in the context of freely migrating cells on 2D substrates are the persistent random motion model~\cite{Gail1970,Selmeczi2005,Selmeczi2008}, L\'evy flights~\cite{Harris2012}, and intermittent dynamics~\cite{Maiuri2015}. We will review these models and their biological implications in section~\ref{sec:single_cell}, and discuss methods for model inference more generally in section~\ref{sec:inference}.
\item Phenomenological models which are constrained in an unbiased and data-driven manner could furthermore yield strong \textbf{constraints for bottom-up models} for the underlying mechanistic basis of the behaviour. These mechanistic models come in different flavours, from minimal mechanical models to active polar gel theories and complex computational implementations. A central difficulty in connecting these models to experiments is that they are frequently under-constrained and over-parameterized. Phenomenological descriptions could provide much more precise `targets' for mechanistic approaches by introducing stronger constraints. Furthermore, they could be used to test conceptual modelling assumptions or approximations, and thus give insight into the key biological processes in a given system. We will discuss this connection in section~\ref{sec:mechanism}.
\item Finally, data-driven frameworks may provide \textbf{systematic frameworks} to address increasingly complex questions, making it possible to add complexity step-by-step. For example, to describe the dynamics of interacting cells, it may be useful to have a theory for the dynamics of single migrating cells. We will discuss how data-driven approaches can help quantify the behavioural variability of migrating cells in section~\ref{sec:variability} and identify models for cell-cell interactions~\ref{sec:collective}.
\end{enumerate}

\section{Learning the stochastic dynamics of single cell migration}
\label{sec:single_cell}

\subsection{An equation of motion for freely migrating cells}
\label{sec:free_migration}

The simplest possible experiment that could teach us something about cell migration behaviour is perhaps the motion of isolated single cells on a uniform two-dimensional (2D) substrate. This is of course not a common setting in physiological processes, in which cells typically encounter heterogeneous, confining three-dimensional (3D) environments $-$ yet it is the archetypal cell migration experiment that has taught us much of what we know about migrating cells. We will turn our attention towards the description of systems that include spatial structures in the next section. Here, we will review what we have learnt from 2D cell migration, and how this may provide a generalizable basis to describe more complex systems.

Even in the simple environment of a uniform 2D substrate, the migration of single cells is non-trivial, as it is powered by a complex cytoskeletal assembly. To study migrating cells, a natural avenue is to focus on the underlying biochemical and biophysical mechanisms, and the molecular pathways underlying them. For this endeavour, the simple scenario of free 2D cell migration was key, and the insights gained have been reviewed elsewhere~\cite{Danuser2013}. An alternative, however, is to zoom out from the molecular level to the behaviour at the cellular scale and to measure the overall motion of the cell. Characterizing these system-level dynamics could then teach us about typical behaviours of cells, which may eventually help understand how such emergent behaviours are generated by the underlying molecular players.

\begin{figure}[h!]
	\includegraphics[width=0.5\textwidth]{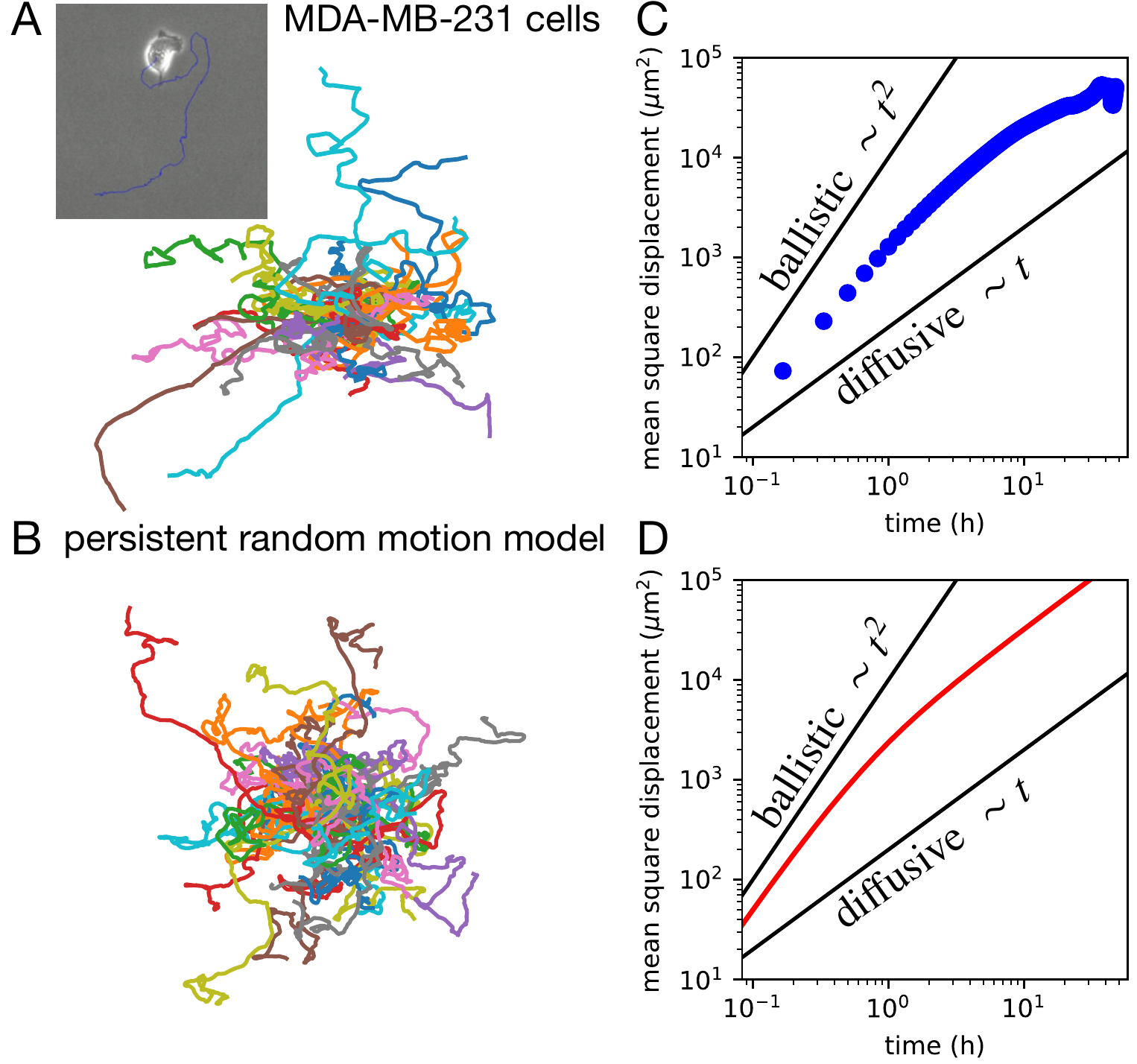}
	\centering
		\caption{
		\textbf{Free 2D cell migration.}
		(A,B) 2D trajectories of single migrating MDA-MB-231 cells and of simulated cells based on the persistent random motion model, respectively. \textit{Inset in A}: Brightfield microscopy image of a migrating MDA-MB-231 cell. 
            (C,D) Mean square displacement curves calculated from experiment and the persistent random motion model (Eq.~\eqref{OU_eom}), respectively. Black lines indicate the limits of ballistic and diffusive motion.
				 }
	\label{fig:2d_migration}
\end{figure}

A simple way to quantify the dynamics of migrating cells is a reduction to a single variable: the position of the cell as a function of time, i.e. the trajectory $\mathbf{x}(t)$ of its nucleus or centroid (Fig.~\ref{fig:2d_migration}A). The first cell tracking experiments were performed over a century ago~\cite{Przibram1917,Furth1920}. At this level, all other putative cellular degrees of freedom, such as the cell shape, cytoskeletal organization, and traction forces, remain unobserved. The trajectory of the cell is thus a minimal representation of a behaviour: it is observed at the cellular scale, and over long time-periods compared to the time-scales of the internal dynamics. The underlying migratory processes give rise to a mix of deterministic trends, visible as persistent segments, and seemingly random, stochastic components. Accordingly, the mean-square-displacement (MSD) exhibits the signatures of ballistic motion at short time-scales and diffusive motion at long time-scales, and is well described by the formula~\cite{Furth1920,Gail1970,Dunn1983} (Fig.~\ref{fig:2d_migration}C,D):
\begin{equation}
\label{MSD}
\langle [\mathbf{x}(t)-\mathbf{x}(0)]^2 \rangle = A (t/\tau_\mathrm{p} + e^{-t/\tau_\mathrm{p}} -1)
\end{equation}
By measuring the MSD, one can therefore recover two key parameters that characterize the behaviour: the persistence time $\tau_\mathrm{p}$, which quantifies the time over which correlations in the cell velocity decay, and the diffusion coefficient $D=A/2d\tau_\mathrm{p}$, where $d$ is the dimensionality. These parameters are frequently used to quantify cell migration, for example to determine the effect of pharmacological treatments of cells, or to contrast different cell types. However, just by measuring the MSD we cannot be sure if these two parameters are sufficient to describe all the statistical features of the observed process. For this, one would need to obtain an equation of motion of the cell that predicts all features of the trajectories, which we discuss next.

The trajectories $\mathbf{x}(t)$ give access to much more information than just the MSD. Specifically, based on the cell trajectories, we can estimate the increments $\Delta \mathbf{x} = \mathbf{x}(t_1)-\mathbf{x}(t_2)$ at various time-scales, including the instantaneous velocities and accelerations of the cell. How should we think about the statistics provided by this additional short time-scale information?

A natural way to think about cell trajectories from a mathematical perspective is the framework of stochastic differential equations, which can provide stochastic equations of motion for migrating cells. A simple model that predicts an MSD of the form of Eq.~\eqref{MSD} is an equation of motion for the cell velocity $\mathbf{v}=\d \mathbf{x}/\d t$, the \textit{persistent random motion model}:
\begin{equation}
\label{OU_eom}
\frac{\d \mathbf{v}}{\d t} = - \frac{1}{\tau_\mathrm{p}} \mathbf{v} + \sigma \boldsymbol{\eta}(t)
\end{equation}
This equation of motion thus predicts the cell acceleration as a function of its velocity, and generates trajectories similar to those observed in experiments (Fig.~\ref{fig:2d_migration}A,B). It consists of two components: a deterministic contribution (first term on the right-hand side), which accounts for the cell persistence, and a Gaussian white noise term (second term on the right-hand side), which accounts for the stochasticity of the motion. This equation predicts the MSD in Eq.~\eqref{MSD} with $A=2\sigma^2 \tau_\mathrm{p}^2$. However, Eq.~\eqref{OU_eom} also predicts many other features of the trajectory dynamics. Specifically, it predicts a Gaussian steady state probability distribution of velocities $p(\mathbf{v})$ with a variance $\tau_\mathrm{p} \sigma^2 /2$, and a velocity auto-correlation function $\langle \mathbf{v}(t) \mathbf{v}(t') \rangle$ that decays as a single exponential with a time-scale $\tau_\mathrm{p}$. Furthermore, it makes a specific prediction about the \emph{conditional average} of the observed cellular accelerations:
\begin{equation}
\label{OU_condav}
\left\langle \frac{\Delta v}{\Delta t} \bigg| v \right\rangle \approx - \frac{1}{\tau_\mathrm{p}} v.
\end{equation}
This conditional average corresponds to the average of the instantaneous acceleration for each observed instantaneous velocity. Note that the derivatives $v=\Delta x/\Delta t$ and $\Delta v/\Delta t$ typically cannot be measured exactly, but are estimated through numerical differentiation of the position trajectories $x(t)$. This leads to non-trivial discretization effects, which we neglect in Eq.~\eqref{OU_condav} and discuss in detail in section~\ref{sec:inference}. These additional statistics beyond the MSD can thus be used to systematically constrain models for 2D cell migration in a data-driven manner. For example, calculating the conditional average on the left-hand side in Eq.~\eqref{OU_condav} can constrain the deterministic term of the description: in principle, the dependence of acceleration on velocity could be non-linear, and this analysis would reveal such an effect in a model-independent manner. Similarly, the magnitude of the stochastic noise term $\sigma$ can be inferred from the variance of the fluctuations in the trajectories.

\begin{figure}[h!]
	\includegraphics[width=\textwidth]{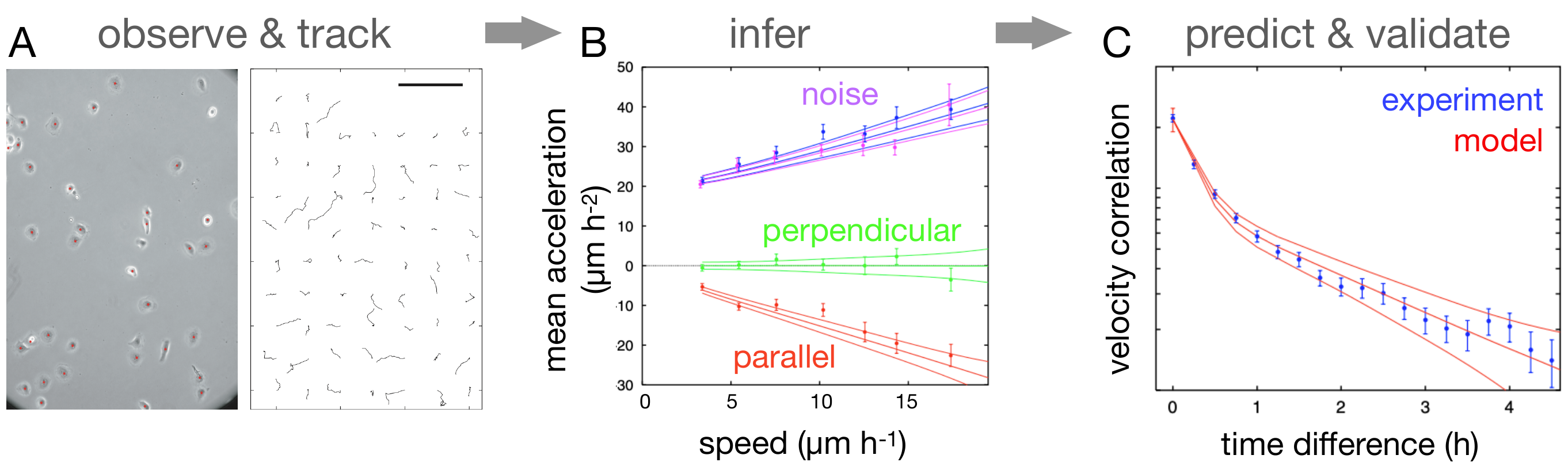}
	\centering
		\caption{
		\textbf{Inference of a dynamical model of 2D cell migration.}
		(A) Observation and tracking step: microscopy image of human dermal keratinocytes (HaCaT) and corresponding nucleus trajectories (scale bar: 200 $\mu$m).
            (B) Inference step: mean and standard deviation as function of speed of the following quantities: average components (calculated using the left hand side of Eq.~\eqref{OU_condav}) parallel (red) and perpendicular (green) to the direction of motion, and stochastic components parallel (blue) and perpendicular (magenta), providing an estimate of $\sigma(v)$. Solid curves show the same quantities, plus/minus one standard deviation, calculated from the inferred model (Eq.~\eqref{memory_kernel}).  
            (C) Prediction and validation step: experimental (blue) and predicted (red) velocity auto-correlation function.
            All panels are adapted from~\cite{Selmeczi2008}.
				 }
	\label{fig:2d_inference}
\end{figure}

As our measurements of cell trajectories have become increasingly accurate and computer-based tracking has allowed generating large sets of such data, a number of statistical features that are not predicted by the persistent random motion model (Eq.~\eqref{OU_eom}) have been identified. Specifically, the velocity distributions of cells are typically not Gaussian, but exhibit exponential tails~\cite{Czirok1998,Cherstvy2018} and the velocity auto-correlation is not exponential, but typically bi-exponential~\cite{Selmeczi2005}. To build a model of free 2D migration that captures these anomalous features, a data-driven approach to learn an equation of cell motion directly from data was proposed by Selmeczi et al.~\cite{Selmeczi2005} (Fig.~\ref{fig:2d_inference}). For this, the conditional average of the acceleration (Eq.~\eqref{OU_condav}) provides a strong constraint on the model (Fig.~\ref{fig:2d_inference}B). Based on this, the authors determined the simplest model consistent with all the observed statistics, which contains an additional memory term in the velocities. We therefore refer to it as the \textit{persistent memory model}. Specifically, the authors identified the following equation of motion based on the data:
\begin{equation}
\label{memory_kernel}
\frac{\mathrm{d} \mathbf{v}}{\mathrm{d} t} = - \beta(v) \mathbf{v} + \alpha^2 \int_{-\infty}^{t} \mathrm{d} t' e^{-\gamma(t-t')}\mathbf{v}(t') + \sigma(v) \boldsymbol{\eta}(t)
\end{equation}
where the multiplicative noise $\sigma(v)$ is interpreted in the It\^{o} sense~\cite{Gardiner1985}. Here, the first term provides a (speed-dependent) time-scale $\tau = \beta^{-1}(v)$ on which the velocity fluctuates around zero, like in the persistent random motion model (Eq.~\ref{OU_eom}). The second term is a memory kernel, which depends on past velocities with a memory time-scale $\gamma^{-1}$. These two time-scales then give rise to a bi-exponential velocity auto-correlation, as observed experimentally (Fig.~\ref{fig:2d_inference}C). Furthermore, this inferred model captures various other anomalous statistics, including the non-Gaussian speed distribution. Similar results were subsequently also found in 2D migration of the amoeba \textit{Dictyostelium}~\cite{Selmeczi2008,Takagi2008,Bodeker2010,Li2011} and breast cancer cells~\cite{Brueckner2019}. Notably, these various studies showed that while the overall form of Eq.~\eqref{memory_kernel} is conserved across cell types, the functions $\beta(v)$ and $\sigma(v)$ had qualitatively different shapes for different cell types. 

Such data-driven, quantitative frameworks for 2D cell migration are useful in several ways. First, it provides a benchmark for characterizing the behaviours of different cell types and determining the effects of drug treatments or genetic perturbations. Secondly, the structure of the inferred model can give insight into the underlying cell dynamics. Importantly, the memory kernel indicates that knowing the current state of motion (determined by the velocity $\mathbf{v}(t)$ at time $t$) is not enough to predict future cell motion, but the history of the process (up to a time-scale given by $\gamma^{-1}$) also needs to be considered. Such memory is presumably encoded in the polar structure of the cell, corresponding to unobserved associated variables that render the dynamics of cell position and velocity non-Markovian (see section~\ref{sec:accelerations}). Importantly, determining Eq.~\eqref{memory_kernel} from the data yields a quantitative description of how these latent variables affect cell motion. Thus, this description can now provide constraints for bottom-up models that seek to connect mechanisms to overall motion. We will discuss this avenue in more detail in section~\ref{sec:mechanism}.

Alternatives to this persistent random memory framework also exist in the literature, including L\'evy walk models for T-cells~\cite{Harris2012,Banigan2015} and fractional diffusion equations~\cite{Dieterich2008} or switching dynamics between modes of movement for epithelial cells~\cite{Potdar2010}. Thus, the type of description may vary depending on the cell type. For example, T-cells and other immune cells have different properties to many of the other epithelial cell types considered here, which could explain the difference in dynamics observed. Furthermore, stochastic equations of motion have been applied to biased cell motion such as chemotaxis~\cite{Tranquillo1987,Tranquillo1988}. Taken together, the minimal example of a freely migrating cell already shows how inferring dynamical models of cell migration can yield insights into the dynamics of living cells beyond simple quantitative readouts.

\subsection{Cell migration as an inference problem}
\label{sec:learning_eoms}

The most remarkable feature of the persistent random motion and the persistent memory models for free 2D cell migration is the drastic reduction in complexity achieved. Small and fast dynamics of the cell contour appear as dynamical noise (not to be confused with technical noise or measurement error), and only a small number of parameters are necessary to accurately capture cell motion at the level of trajectories. The data-driven development of these models therefore formalizes concepts such as persistence and cellular fluctuations. Indeed, an important step in the inference procedure was to disentangle the deterministic (average) and stochastic (fluctuating) components of the dynamics. Decomposing these two contributions is a key advantage of learning the stochastic equation of motion of the system, and then allows interpretation of each component.

While the persistent random motion framework is intuitive and is frequently used to describe cell migration, the aim of the approach outlined in the previous section was to determine a dynamical model for single cell migration without such prior intuition, directly from data. More specifically, the aim was to learn an equation of motion from the stochastic cell trajectories. This places this work into a general class of inverse problems where the aim is to derive a physical description from data in an unbiased manner. This inference principle is a key technique whose full power becomes apparent when used with rigorous inference methods and on complex data sets. Here, by rigorous inference, we mean inference techniques that provably converge to the correct result for simulated data sets. We are by no means constrained to infer cell acceleration as a function of velocity: what if the migration takes place in a complex structured environment? Then, other degrees of freedom, such as the cell position, can be used as conditioning variables. We can therefore infer how cellular responses (measured in accelerations) depend upon the local geometry or structure of the environment (measured by position). We will discuss such an approach in section~\ref{sec:confined}. Furthermore, we can imagine tracking other degrees of freedom of the cell beyond its position, for example protrusions and retractions, or even spatially extended variables such as shape or internal concentration fields. Deriving the equations of motion of these degrees of freedom, and their coupling to each other and to the environment, could yield key insights into cell behaviour. This approach could provide more direct connections with mechanistic models (see section~\ref{sec:mechanism}). Finally, new inference techniques also allow for inference in high-dimensional and interacting systems~\cite{Frishman2018,Brueckner2020a}, which could be used to learn the dynamics of interacting cells in collective migration (section~\ref{sec:collective}). The data-driven persistent random motion framework introduced in the previous section establishes a conceptual basis to understand these other approaches, which become increasingly complex when we go beyond this simple stochastic process. Inferring an equation of cell motion based on experimental trajectories has helped to elevate persistent random cell motion from a concept into a theory, meaning that we progress from a somewhat fuzzy intuition to a mathematical equation that makes falsifiable predictions that can be tested on the data. We will highlight avenues for achieving something similar for more complex systems. For this, we will first turn to the example of a single cell migrating in a standardized structured environment, allowing inference of its interaction with external features. To enable going through such an example in detail, we will discuss a biased selection of the literature and focus on our work of learning the equation of motion of a cell confined in a two-state micropattern~\cite{Brueckner2019}. In the following sections, we will then discuss the much broader literature on inferring cell-to-cell variability, connecting to bottom-up models, and learning models of collective migration.

\subsection{Cell migration in structured environments}
\label{sec:confined}

Cell migration on unstructured 2D substrates provides an important benchmark for how to think about cell migration dynamics, and its simplicity has allowed significant theoretical progress. However, in physiological processes, cells do not encounter such unstructured environments: they navigate extra-cellular environments that are complex, structured, and confining. These include collagen matrices, bone marrow, or blood vessel linings~\cite{Friedl2003}. Thus, if we want to understand cellular dynamics in physiological processes, we need to study confined cell migration. Cell migration in 3D extra-cellular matrices has been studied extensively (see reviews in refs.~\cite{Even-Ram2005,Driscoll2015a,Vargas2017}). However, these matrices are spatially heterogeneous, and thus single cells will only rarely encounter the same obstacle twice. While some studies have made progress on quantifying cell trajectories through ECM~\cite{Wu2014} as well as bacterial motion through heterogeneous porous media~\cite{Bhattacharjee2019,Bhattacharjee2019a}, it is in general difficult to gather sufficient statistics to understand how the local microstructure determines the cell behaviour. A popular approach to study confined migration while keeping the extra-cellular environment as simple as possible, are \textit{in vitro} artificial confining geometries. Such geometrical confinements can be implemented using micropatterning, 3D printing, or microfluidics, and can be designed to expose cells to challenges such as overcoming a constriction or navigating a maze. Overcoming such challenges is an inherent feature in \textit{in vivo} contexts, and is clearly an aspect that is missed by studying cells in featureless 2D surfaces. In this section, we will first discuss the key experimental approaches to study confined cells \textit{in vitro}, before turning to inference approaches for confined migration.

\begin{figure}[h!]
	\includegraphics[width=0.9\textwidth]{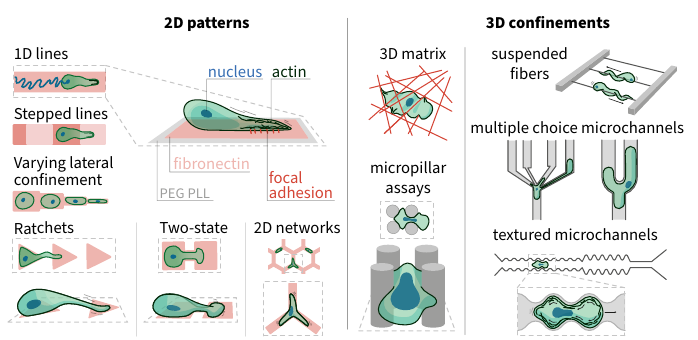}
	\centering
		\caption{
		\textbf{Experimental approaches for studying confined cell migration.}
		2D confining geometries of cells are typically designed using micropatterning, in which a region of defined geometry is coated with a cell-adhesive protein, fibronectin, while the surroundings are passivated with cell-repellent PEG-PLL polymers. To study cell migration, such micropatterns have been used in the shape of 1D lines, stepped lines with varying protein coating density, varying lateral confinement, series of triangles in a ratchet-like arrangement, two-state micropatterns, and 2D networks of 1D lines, such as hexagonal networks. 3D confinements to study cell migration include 3D extracellular matrices, micropillar arrays, suspended fibers, multiple-choice microchannels, as well as textured microchannels. In all these systems, there is not only basal, but also lateral confinement, causing among other things deformation of the cell nucleus when cells migrate through constrictions.
				 }
	\label{fig:confined}
\end{figure}

Artificial systems to study confined migration include 2D micropatterns~\cite{Singhvi1994,Chen1998}, microfluidic devices~\cite{Lin2015}, 3D confinements~\cite{Renkawitz2019,Kopf2019,Reversat2020}, micropillar arrays~\cite{Davidson2020}, and suspended nanofibers~\cite{Zhang2007,Guetta-Terrier2015,Singh2021} (Fig.~\ref{fig:confined}). These systems allow monitoring of large numbers of cells migrating in identical, standardized structured environments, yielding unprecedented large data sets on cell behaviour. Micropatterning provides a simple way to confine cells: using differential surface coatings, one can define areas to which cells can adhere, surrounded by cell-repellent regions. With this technique, confinements of arbitrary geometrical shape can be produced, giving access to a wide variety of systems. One of the simplest migration experiments using micropatterns is confinement to narrow stripes~\cite{Maiuri2012}. In such effectively one-dimensional (1D) confinements, cells typically perform persistent random motion in one dimension~\cite{Fraley2012}. This 1D mode of migration has been proposed as a model for aspects of cell migration in 3D extra-cellular matrices: in 3D matrices, cells frequently encounter narrow channels through which they migrate, reminiscent of an effective 1D confinement~\cite{Doyle2009,Fraley2012,Guetta-Terrier2015}. Indeed, the morphology of cells on narrow 1D lines is highly stretched, similar to morphologies observed in 3D, which do not feature the broad fan-like lamellipodia observed on 2D substrates~\cite{Doyle2009,Chang2013,Guetta-Terrier2015}. However, unlike 1D lines, physiological extra-cellular environments are structured, for example through the presence of thin constrictions through which cells need to squeeze during migration~\cite{Park2015,Green2018,Paul2017,Wolf2013}. To study the response to such constrictions, micropatterned lines with periodic modulations, or gaps, which cells need to overcome have been developed. For example, ratchet-like confinement geometries were found to rectify the direction of motion of cells~\cite{Mahmud2009,Caballero2014,Comelles2014,LoVecchio2020}, a process termed ratchetaxis (see ref.~\cite{Caballero2015} for a review). Using a microfluidic confinement with walls featuring similar ratchet-like modulations, a novel mode of migration relying on friction with the local topography of the walls was revealed~\cite{Reversat2020}. Increasing the complexity of the environments even more, experimental systems have been developed to study how cells make decisions at junctions featuring either two symmetric~\cite{Hadjitheodorou2022} or several constrictions of varying widths~\cite{Renkawitz2019,Kopf2019}, which revealed the intra-cellular processes involved in cellular decision making in such systems. Finally, another approach to study cells overcoming constrictions is to consider geometries where the boundaries on both sides are closed, meaning that the cell has to turn around and make transitions back and forth across the same constriction. This was done using two-state micropatterns, which have the advantage that long trajectories of subsequent transitions can be obtained~\cite{Brueckner2019,Fink2019}. 

These experimental approaches using standardized confinements have given insight into intra-cellular processes~\cite{Renkawitz2019,Kopf2019} and have yielded quantitative cellular readouts, for example the degree of directionality in ratchetaxis~\cite{LoVecchio2020}, switching rates between run and rest states on 1D lines~\cite{Schreiber2016}, or transition rates in two-state micropatterns as a function of the geometry~\cite{Brueckner2019,Fink2019}. Based on our discussion of free 2D cell migration, a key challenge to go beyond cellular readouts from confined migration experiments is to develop an equation of cell motion that accounts for structured environments. In this case, the terms of the equation of motion will depend on both the position and velocity of the cell. As these cells solve the challenge of navigating their confining environment, the terms of the equation of motion give insight into how cells dynamically solve this problem and thus encode how it responds to the structures in its environment, which we will discuss in the next section.

\subsection{Learning an equation of confined cell motion}
\label{sec:two_state}
\begin{figure}[h!]
	\includegraphics[width=\textwidth]{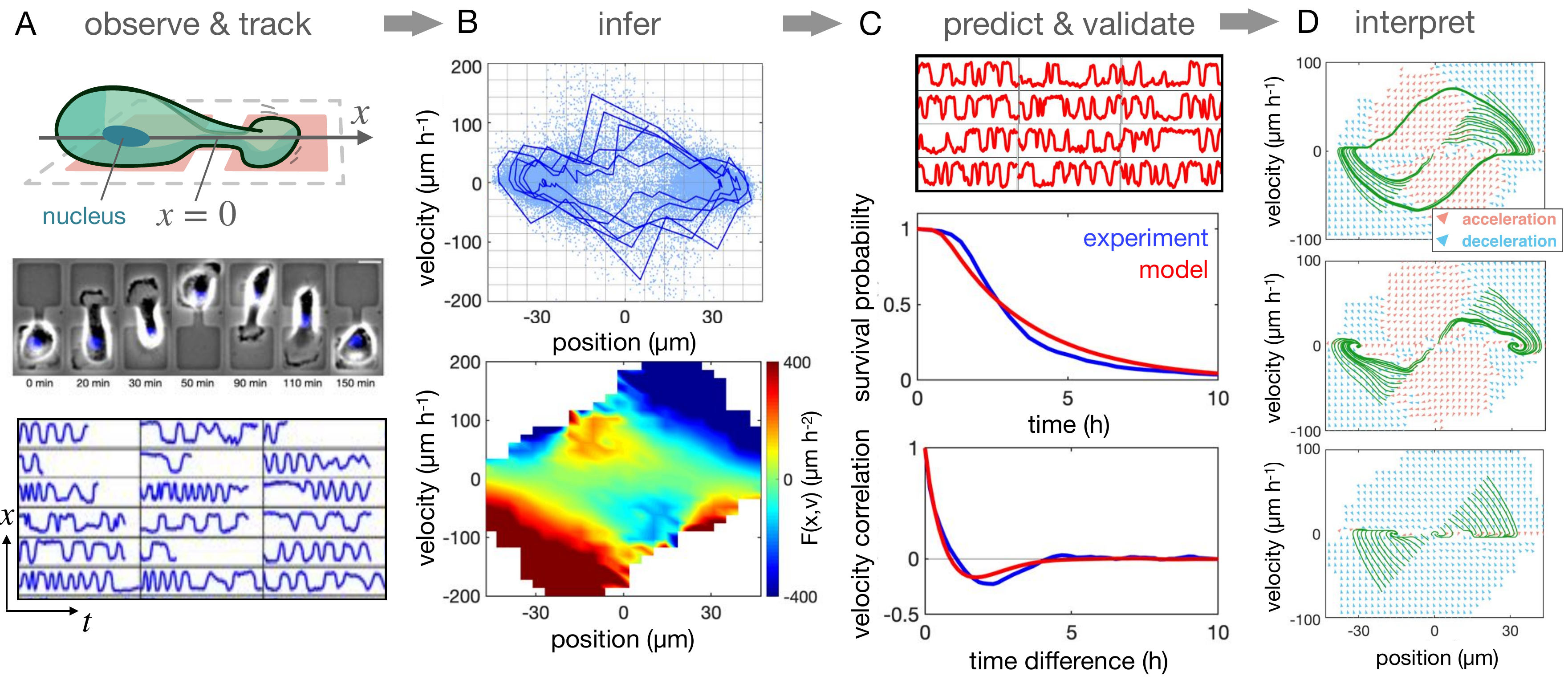}
	\centering
		\caption{
		\textbf{Inferring an equation of confined cell motion.}
            (A) Observation and tracking step: human breast cancer cells (MDA-MB-231) are confined to two-state micropatterns and imaged at 10 min time intervals (scale bar: 25 $\mu$m). Bottom: nucleus trajectories as a function of time, plotted for a (0,50h) interval.
            (B) Inference step: single-cell trajectory in $xv$-space (blue line) and recorded data points from a large data-set of cells (lightblue points, top). Averaged together, this gives the deterministic term $F(x,v)$ (bottom). 
            (C) Prediction and validation step: trajectories predicted based on the inferred model (top); experimental (blue) and predicted (red) survival probability $S(t)$, measuring the probability that a transition across $x=0$ has not occured after time $t$ (middle); normalized velocity auto-correlation functions (bottom)
            (D) Interpretation of the inference results: trajectories (green) of the deterministic dynamics for a number of different initial conditions. The flow field is shown by arrowheads, where acceleration is orange and deceleration is blue; shown for MDA-MB-231 cells (top), MCF10A cells (middle); and MDA-MB-231 cells migrating in a system without constriction.
				 }
	\label{fig:two_state_inference}
\end{figure}

Learning a data-driven model of confined cell migration requires large data sets of trajectories, which can be obtained using minimal \textit{in vitro} confinements. In previous work, we used two-state micropatterns as a minimal system to study how cells overcome thin constrictions in confining environments~\cite{Brueckner2019}. To provide a pedagogical example of how one can learn an equation of motion from confined cell migration data, we will discuss this example here in more detail. These micropatterns consist of two square adhesive islands connected by a thin adhesive bridge (Fig.~\ref{fig:two_state_inference}A). This setup leads to repeated stochastic transitions of the cells between these two islands, with large variability both over time and across cells. Based on the trajectories of these cells, we then developed a generalization of the persistent random motion model (Eq.~\eqref{OU_eom}) to the problem of confined migration. An important assumption in  the persistent random motion model is the uniformity and isotropicity of space: the cellular dynamics are assumed to be independent of position, and the same in all directions. Clearly, these assumptions are no longer valid in structured systems. This suggests a more general formulation of an equation of cell motion for confined migration, in which the dynamics can also depend on the position $x$ of the cell, which we refer to as an \textit{equation of confined cell motion}:
\begin{equation}
\label{Fxv_eom}
\frac{\d v}{\d t} = F(x,v) +\sigma(x,v) \eta(t)
\end{equation}
where $F(x,v)$ is a generalized version of the deterministic term in Eq.~\eqref{OU_eom}, and $\sigma(x,v)$ is the amplitude of the stochastic fluctuations. Note that in the presence of state-dependent noise, meaning that $\sigma(x,v)$ is not a constant, the inferred deterministic term depends on the chosen noise-convention~\cite{Gardiner1985}. Here and throughout the text, this equation is interpreted in the It\^{o} sense, but note that the inferred deterministic term $F$ would differ in the Stratonovich convention if the noise is $v$-dependent. Put simply, $F(x,v)$ is the average acceleration of the cell as a function of its position $x$ and its velocity $v$.  Importantly, other descriptions for the dynamics are in principle possible, and this postulated equation could be incorrect. Thus, once a model of this form has been inferred, one has to test its predictive power and contrast it with that of alternative descriptions, which we discuss below. Note that in this case, the dynamical description is one-dimensional, as the lateral dimensions are highly constrained by the pattern. Furthermore, we here start with a memory-less description, which is simpler than the memory kernel equation of motion for 2D migration (Eq.~\eqref{memory_kernel}). Thus, the inference procedure starts with the simplest model which is only modified when the data demands it. The aim is now to determine the structure of the dynamical terms $F$ and $\sigma$ in a completely data-driven method based on the experimental trajectories. Specifically, to a first approximation, the deterministic term of this equation can be inferred using a conditional average of the observed cellular accelerations:
\begin{equation}
\label{Fxv_condav}
F(x,v) \approx \left\langle \frac{\Delta v}{\Delta t} \bigg| x,v \right\rangle
\end{equation}
which is the generalized formulation of Eq.~\eqref{OU_condav} for an equation of motion with positional dependence. This approach works as follows: the trajectories are represented in the position-velocity phase space, which is split into bins using a regular grid (Fig.~\ref{fig:two_state_inference}B, top). In each bin, the average acceleration is measured (Eq.~\eqref{Fxv_condav}), giving the deterministic term $F(x,v)$ (Fig.~\ref{fig:two_state_inference}B, bottom). Similarly, but calculating the standard deviation of fluctuations, the stochastic term $\sigma(x,v)$ can be inferred. Note that this binning approach relies on an approximate estimator that contains bias terms due to the numerical derivatives used to obtain velocities and accelerations. This can be corrected with a more data-efficient approach relying on a set of smooth basis function such as polynomials or Fourier components, which we discuss in section~\ref{sec:ULI}.

Importantly, while the experimental data is used to constrain the shape and parameters of the deterministic dynamics $F(x,v)$, there is no guarantee that this approach yields an adequate representation of the dynamics of the system over a broad range of time-scales: the inference approach relies on the assumption that the dynamics of the system can in fact be described by the equation of motion Eq.~\eqref{Fxv_eom}, which could fail in many ways. 

On the one hand, the dynamics could be more complex and could require additional memory terms~\cite{Selmeczi2005}, a time-dependent description~\cite{Metzner2015}, or an explicit description of the cell-to-cell variability~\cite{Wu2014}. To test the validity of this description, we therefore need to perform a test of predictive power. Specifically, to perform the inference, we constrained the equation of motion solely based on the short time-scale information provided by the experimental trajectories, including the velocities and accelerations of the cell. Thus, as an independent test of the model~\cite{Selmeczi2005,Brueckner2019}, we predict statistics quantifying the cell behaviour on long time-scales, for example the distribution of transition times or the velocity auto-correlation function, which all capture the experimentally observed statistics (Fig.~\ref{fig:two_state_inference}C). 

On the other hand, the dynamics could also be simpler and we have to ensure that we identified the simplest model consistent with the data. To address this, we increased the complexity of inferred models step-by-step and ruled out the possible simpler models. For instance, an alternative inference based on a first order equation of motion (for $\d x / \d t$ as opposed to $\d v / \d t$) was unable to capture the data. Furthermore, simplifications of the general non-linear term $F(x,v)$ into a separable form $F(x,v)=F_x(x)+F_v(v)$, as would be the case for a conservative potential $V(x)$, such that $F_x(x)=-\partial_x V(x)$, were inconsistent with the data. Based on this, we concluded that Eq.~\eqref{Fxv_eom} was the simplest model that could capture the data. These examples already show how exploring models that do not describe the data can be very instructive, as they allow to rule out simple hypotheses.

In the example of the confined cell problem, we found that an insightful representation of the system can be achieved by examining the deterministic dynamics of the system in a phase-portrait of position and velocity (Fig.~\ref{fig:two_state_inference}D). Intuitively, one might expect that the hopping behaviour across the thin constriction placed by the micropattern could be generated by a noisy cellular activity competing with an effective energy barrier placed by the constriction. Strikingly, however, the inferred map of the deterministic accelerations reveals that cells have a tendency to accelerate into the constriction. In fact, the flow field of the deterministic dynamics exhibits an excitable flow, where a small noise-driven perturbation leads to a large excursion in the phase space due to a deterministic amplification of the cell speed. This amplification is observed in both cancerous (MDA-MB-231) and non-cancerous (MCF10A) cells, suggesting that it may be a generic cellular response to thin constrictions. Indeed, in systems in which the constriction is removed, the amplification vanishes (Fig.~\ref{fig:two_state_inference}D, bottom). This approach also reveals that the non-linear dynamics are poised close to a bifurcation between a limit cycle and a bistable system. Interestingly, different cell lines exhibit behaviours on both sides of this transition: MDA-MB-231 cells exhibit a limit cycle, while MCF10A cells show excitable bistable dynamics. Thus, the deterministic phase-portrait implies that the cancerous cells have a stronger tendency to overcome the constriction, while the non-cancerous cells rely on stochastic fluctuations to perform transitions. 

In the next section, we will discuss how we can use these insights to quantify and characterize the striking variability in the observed cell behaviours, which are already apparent at the level of the cell trajectories. Moreover, this approach could help advance our understanding of locomotion at the molecular level by providing constraints for bottom-up models that connect microscopic rules to the system-level dynamics of cells. Finally, the insights gained based on this framework could provide a generalizable basis to investigate the dynamics of assemblies of interacting cells. We will discuss both of these aspects in the following sections.

\subsection{Why are cell migration dynamics underdamped?}
\label{sec:accelerations}

The equation for 2D persistent random motion (Eq.~\eqref{OU_eom}) and the equation of motion for confined cell migration (Eq.~\eqref{Fxv_eom}) share a key feature: both are stochastic differential equations that are second-order in time, and therefore a manifestation of the \emph{underdamped} Langevin equation. These equations predict the acceleration as a function of position and velocity. This is in contrast to first-order stochastic equations of motion which are frequently used to describe the motion of overdamped Brownian systems subject to thermal noise~\cite{Frey2005}. For such overdamped Brownian systems, the effects of inertia can be neglected at time-scales larger than the velocity relaxation time $m/\zeta$, where $\zeta$ is the friction coefficient and $m$ is the mass of the particle. Therefore, friction is directly equated with the sum of thermal and external forces, yielding a first-order, \emph{overdamped} Langevin equation. However, the same physical argument applies to migrating cells: the forces acting on cells, including frictional forces, are much larger than the inertial term $m \dot{v}$, and thus we can take $m \approx 0$ to a very good approximation. Why then are cell migration dynamics described by underdamped equations of motion? 

An underdamped equation describes a process in which velocities have temporal correlations, and do not just follow a white noise process as in overdamped systems. Physical inertia is one way of introducing temporal correlations, as the inertia prohibits instantaneous reversals of direction, and instead introduces a characteristic time scale to adjust velocities. Similarly, cells do not instantaneously change their direction if they are in a polarized state, meaning that polarization gives rise to a kind of "effective inertia". To be precise, the cell's propulsive forces constitute a stochastic process with correlation time-scales similar to the migration time-scales, and therefore introduce correlations in the cell velocities. 

This idea can be demonstrated with a very simple model of the overdamped dynamics of a confined migrating cell that is driven by a self-propulsive cell polarity $P(t)$,
\begin{align}
\label{eq_overdamped_x}
\dot{x} &= f(x) + P(t) \\
\label{eq_overdamped_p}
\dot{P} &= g(x,P) + \sigma \eta(t)
\end{align}
where $f(x)$ are the forces acting on the cell in a confining environment, and $g(x,P)$ is a general formulation of polarity dynamics that may depend on both the current polarity and the position of the cell. Here, $P(t)$ subsumes all of the subcellular processes mentioned above that determine the direction of self-propulsion of the cell. Then, taking the derivative of Eq.~\eqref{eq_overdamped_x} and substituting Eq.~\eqref{eq_overdamped_p}, we obtain:
\begin{equation}
\label{eq_overdamped_Fxv}
\dot{v} = \underbrace{ f'(x)v + g(x,v-f(x)) }_{F(x,v)} + \sigma \eta(t)
\end{equation}
This shows how an overdamped particle that is driven by underlying time-correlated polarity dynamics exhibits effective underdamped stochastic dynamics. The deterministic term $F(x,v)$ is determined by a non-trivial combination of the confinement forces $f(x)$ acting on the cell and the polarity dynamics $g(x,P)$. Importantly, this also means that we should not think of the deterministic term $F(x,v)$ in Eq.~\eqref{Fxv_eom} (and equivalently the term $-\tau_\mathrm{p}^{-1} \mathbf{v}$ in Eq.~\eqref{OU_eom}) as physical force fields, but as an acceleration field that is determined by the underlying time-correlated machinery of the cell.

The underlying molecular processes that determine the cell polarity $P(t)$ are complex, but can be understood as an interplay of actin flows and various polarity-mediating molecular factors. Importantly, these propulsive forces should not be confused with the traction forces exerted by the cell onto the substrate. Indeed, cellular tractions are typically much larger than the forces needed to migrate~\cite{Oliver1999,Elosegui-Artola2018a}. For instance, in keratocytes, traction forces are up to tens of nN~\cite{Lee1994}, while the propulsive force of the leading edge was recently measured to be of the order of 1 nN~\cite{Mohammed2019a}. Instead, the polarity is related to the intracellular concentrations of polarity cues and the actin flows, together determining the cell speed. Specifically, for a given actin polymerization rate, the speed of a migrating cell is determined by the retrograde flow of actin, which is being polymerized at the leading edge, and depolymerized at the trailing edge: the slower the flow, the faster the cell~\cite{Lee1993,Maiuri2015}. Note, however, that slower retrograde flow leads to higher traction, and thus there is an indirect correlation between traction force magnitude and cell speed~\cite{Elosegui-Artola2018a}. The directionality of the actin flow is in turn determined by the concentration profiles of internal signalling cues within the cell, which reorient on long time-scales~\cite{CallanJones2016} (described in our example by Eq.~\eqref{eq_overdamped_p}). Reorientations of these polarity fields lead to changes of the cell velocity vector, i.e. accelerations. Therefore, cellular accelerations are changes of velocity that are determined by intra-cellular dynamics, and not by a net force acting on the cell. Therefore, to understand the origin of the emergent cell migration dynamics, quantified by $F(x,v)$, we should consider how internal degrees of freedom of the cell, including the cell shape, protrusion formation, and polarity determine the net movement of the cell, and how these degrees of freedom couple to the external environment. 

Contrasting the overdamped formulation (Eqs.~\eqref{eq_overdamped_x},~\eqref{eq_overdamped_p}) with the underdamped one (Eq.~\eqref{eq_overdamped_Fxv}) suggests an important conceptual insight into how inferred cell migration models can be connected to more mechanistically interpretable models. Clearly, the overdamped dynamics are physically more interpretable, as they connect directly to the known physics of self-propelled active particles~\cite{Romanczuk2012a} and the individual terms have a physical interpretation. However, inferring such overdamped equations for position and polarity from experimental data is currently an open challenge. To infer such equations from data, one would need trajectories of the cell polarity $P(t)$. However, there is no unique molecular marker of cell polarity, and for candidate markers of cell polarity, such as Rho GTPase localization, it is experimentally challenging to collect large data sets of cell migration trajectories with motion and polarity tracked simultaneously (see ref.~\cite{DAlessandro2021} and section~\ref{sec:beyond_traj} for a more detailed discussion). In contrast, the underdamped formulation requires only tracking of the cell nucleus, from which the velocity degree of freedom can be obtained through numerical differentiation. Thus, learning the underdamped dynamics of migrating cells from data can provide a key step towards understanding more mechanistic aspects. Indeed, the mapping from overdamped to underdamped dynamics suggested by Eq.~\eqref{eq_overdamped_Fxv} could provide a way to link mechanistic and inferred models more directly, by comparing the predicted $F(x,v)$ of postulated active particle models to the inferred underdamped equation of motion.

\section{Learning equations of motion from stochastic trajectories}
\label{sec:inference}

In the previous section, we discussed how inferring equations of cell motion gives insight into free and confined cell migration. In this section, we  discuss the technical aspects of performing stochastic inference. Please note that this section is not essential to understand the remainder of the review, and can therefore be skipped. Inferring equations from experimental data is a general problem that has been applied to a broad variety of physical and biological systems, ranging from dust particles in a plasma~\cite{Yu2022} to protein diffusion~\cite{Beheiry2015,PerezGarcia2018}, animal~\cite{Stephens2008,Stephens2011} and robotic~\cite{Boudet2021} behaviour, and neural dynamics~\cite{Genkin2020}. There is a long history of inferring dynamical systems from trajectories of deterministic systems~\cite{Crutchfield1987,Daniels2015,Brunton2015,Champion2019,Chen2021}. Such inverse problems are notoriously harder in stochastic systems such as migrating cells: it requires disentangling the stochastic from the deterministic contributions, both of which contribute to shape the trajectory. Importantly, however, fluctuations can also help to make a data set more informative about the system: in low-noise systems, the trajectory may only sample a very narrow region of the phase space, making it difficult to estimate the underlying dynamical system. Thus, successful inference typically requires a data set with sufficient diversity, which may pose a problem in highly stereotyped behaviours such as in morphogenesis. 

A number of methods are now available to perform such equation inference in stochastic overdamped (first-order) equations~\cite{Siegert1998a,Ragwitz2001,Beheiry2015,PerezGarcia2018,Frishman2018,Boninsegna2018,Dai2020a,Callaham2021,Huang2022} as well as underdamped (second-order) systems~\cite{Brueckner2020a,Ferretti2020}. Note that in addition to dealing with the intrinsic stochasticity of the system, realistic experimental data sets are also invariably subject to measurement error, which can have a major impact on numerical derivatives, and requires specialized estimators that are robust to such errors~\cite{Frishman2018,Brueckner2020a}. In this section, we will first lay out the general principles of stochastic inference. Then, we focus on the specific case of performing inference of underdamped equations of motion which is relevant to cell migration trajectories.

\subsection{General principles}
\label{sec:inference_principles}
The overarching idea of equation of motion inference from a complex biological system is to derive a simple physical description of a small number of degrees of freedom (DOFs) that does not require knowledge of all the microscopic details of the system. Thus, the idea is to identify the important DOFs that may follow relatively simple dynamics, that are slow compared to the time-scales of the microscopic processes. Developing an equation of motion model from experimental data in general involves five key steps, which were already illustrated in Figs.~\ref{fig:2d_inference} and~\ref{fig:two_state_inference} using the examples of free and confined cell migration, respectively. Here, we will discuss these steps in a more general context, and illustrate them with the example case of underdamped equations of motion, as these are used to describe cell trajectories (see section~\ref{sec:accelerations}), although the key points are equally relevant for overdamped stochastic systems~\cite{Siegert1998a,Ragwitz2001,Beheiry2015,PerezGarcia2018,Frishman2018}.

\textbf{(1) Observation}: in the first step, the important DOFs of the system have to be identified and observed. These DOFs have to be experimentally accessible and trackable over time to yield the trajectories $x(t)$. Furthermore, to enable inference and interpretation of the model, this set of DOFs should ideally be low-dimensional and therefore provide a minimal representation of the system. In general, there is no principle to determine which DOFs should be tracked, and to some degree it is a choice that is made based on intuition and technical feasibility. The key objective is to arrive at a set of DOFs that allow construction of a predictive model (see point 3). In the examples of free and confined cell migration, this was done by simply measuring the trajectories of the cell nucleus (Figs.~\ref{fig:2d_inference}A, \ref{fig:two_state_inference}A). Identifying the relevant DOFs become even more challenging in collective multicellular settings, as discussed in section~\ref{sec:collective}. In general, if the inference procedure proves to be difficult in the later steps, a different set of DOFs may need to be chosen.

\textbf{(2) Inference}: the second step is the inference of a model from the observed trajectories. In this step, a general, unbiased formulation of a stochastic dynamical system for the tracked DOFs should be postulated, which can then be systematically constrained using the data. To go from the data all the way to the inferred equation, three key steps need to be considered: \\
\textit{(2.1) Equation selection}: the first step is to select the structure of the equation of motion to be inferred from the data. In practise, this selection can often be done based on physical intuition. More principled approaches include searching for maximum predictability from delay embeddings~\cite{Ahamed2021}, testing of Markovianity from data~\cite{Friedrich1998}, or determining the scaling of increments with time~\cite{Vestergaard2015}. For cell migration experiments where the polarity remains unobserved, the appropriate equations are typically underdamped equations of motion for the dynamics of the cell velocity (see section~\ref{sec:accelerations}). \\
\textit{(2.2) Basis selection}: to infer the equation of motion an appropriate representation of the dynamical terms must be chosen. In the confined cell example, this corresponds to choosing how to approximate the functions $F(x,v)$ and $\sigma(x,v)$ by a set of basis functions. In this case, the dynamical terms are represented as a truncated basis expansion
	\begin{equation}
	\label{expansion}
	F(x,v) \approx \sum_{\alpha=0}^{N_b} F_{\alpha} c_\alpha(x,v)
	\end{equation}
where $\{ c_\alpha(x,v) \}$ is the set of basis functions and $N_b$ is the number of functions. Note that this expansion is written for a one-dimensional system, but all expressions generalize straightforwardly to multidimensional systems~\cite{Brueckner2020a}. Thus, the problem of inferring the equation of motion is reduced to estimating the parameters $F_{\alpha}$. If the noise is state-dependent, a similar expression can be written for the stochastic term $\sigma^2(x,v)$. The key problem is then to select the set of basis functions $\{ c_\alpha(x,v) \}$ that is appropriate for the problem at hand. These can be constrained by taking into account the symmetries of the system~\cite{Frishman2018,Brueckner2020}, by applying Bayesian approaches to the fit-complexity trade-off~\cite{Guimera2020}, or by applying sparsity constraints to detect the relevant terms, such as SINDy~\cite{Brunton2015,Champion2019}, which has recently been generalized to stochastic systems~\cite{Boninsegna2018,Dai2020a,Callaham2021,Huang2022}. Based on such principles, one can then determine, for instance, if the noise in the system is state-dependent, or if it should be fitted by a constant amplitude. \\
\textit{(2.3) Estimators}: finally, to perform the inference of the parameters specified by the selected basis in a rigorous manner, one must use the correct estimators to determine these parameters from the observed trajectories. This ensures that the inferred parameters converge to the correct result for simulated data sets with known parameters. This is challenging due to the stochasticity of the system, the inevitable discreteness of the sampled trajectories, and the presence of measurement errors (see next part of this section). These estimators then allow inference of the dynamical terms (Fig.~\ref{fig:2d_inference}B, \ref{fig:two_state_inference}B). 

\textbf{(3) Self-consistency}: before testing the predictive power of the model, there are two tests of self-consistency that should be performed. \\
\textit{(3.1) Noise correlation}: a key assumption of stochastic inference approaches is that deterministic and stochastic contributions can be separated, which relies on the assumption of white noise of the stochastic term $\eta(t)$, such as in Eq.~\eqref{Fxv_eom}, meaning that $\eta(t)$ is uncorrelated in time, $\langle \eta(t) \eta(t') \rangle = \delta(t-t')$. To test the self-consistenty of this assumption, one can calculate the trajectories of the noise increments $\Delta W(t) = \int_t^{t+\Delta t} \eta(s) \ \mathrm{d} s$. Specifically, an empirical estimator for $\Delta W(t)$ is~\cite{Selmeczi2005,Stephens2008,Brueckner2019,Brueckner2021}:
\begin{equation}
\Delta W (t) \approx \frac{\Delta t}{\sigma} \left[ \dot{v}(t) - F(x(t),v(t)) \right]
\end{equation}  
where $F$ and $\sigma$ are the inferred inferred deterministic and stochastic terms, respectively. Then, the auto-correlation $\langle \Delta W (t) \Delta W(t') \rangle$ can be calculated, which should decay to zero within a single time-step if the white noise assumption was correct. Note that at the first time-step $|t-t'|=\Delta t$, a weak negative correlation can appear due to the presence of measurement errors~\cite{Pedersen2016,Brueckner2019}. If this criterion is not satistified, one typically has to revisit point 2.1 and consider a different class of models. For example, if a first-order equation inference is applied to cell migration trajectories, long time-scale noise correlations will appear, since a second-order equation is required. \\
\textit{(3.2) Re-inference}: a second criterion for a self-consistent model is that when new trajectories are simulated based on the inferred model, applying the same inference procedure to these simulated trajectories should yield a consistent result~\cite{Brueckner2020,Brueckner2021}. An important aspect of this is that inferred equation can turn out to be unstable, meaning they fit the data locally in time, but diverge for long time intervals. This needs to be checked and avoided. If these criteria is not satistified, it is likely that points 2.2 or 2.3 should be revisited, or an insufficient amount of data was used in the inference.

\textbf{(4) Validation and prediction}: steps 1 and 2 of the inference make assumptions about the system that could be incorrect, and thus the predictive power of the inferred model must be tested to validate it. The stochastic inference approaches described in section~\ref{sec:two_state} and~\ref{sec:ULI} use as input only the short time-scale information of the trajectories, through the increments corresponding to velocities and accelerations. A key test of the model is then to predict long time-scale statistics that were not used in the inference. Which statistics are suitable for such prediction depends on the system at hand. For instance, for both free and confined migration the velocity auto-correlation function was a natural prediction target (Fig.~\ref{fig:2d_inference}C, \ref{fig:two_state_inference}C). If the predictive power of the model is low, one typically has to revisit points 2.1-2.3 to consider if the correct equation, a reasonable basis, and valid estimators have been used. Note that depending on the method and the amount of hyperparameter tuning in the inference, one may also want to consider splitting the data into training, validation and testing data sets, using standard approaches to such issues. Finally, to further challenge the model, one may want to test its predictive power on other experiments not used for training, such as mutants, perturbations, or other environmental conditions (such as a different micropattern geometry).

\textbf{(5) Interpretation}: having determined a valid model for the observed dynamics, this model can be interpreted to gain insight into the system. This last step is of course very much system-dependent. An important aspect of the stochastic inference approach is the decomposition of the dynamics into deterministic and stochastic components, i.e. $F$ and $\sigma$. Based on this decomposition, these components can be interpreted separately, and their respective contributions to the dynamics can be conceptualized. For example, this decomposition revealed distinct classes of non-linear dynamical systems in the position-velocity phase space of confined migrating cells (Fig.~\ref{fig:two_state_inference}).

\subsection{Sources of inference error}
\label{sec:ULI}

When following the sequence of steps laid out in the previous section, there are multiple sources of inference error, which can lead to deviation of the observed and the predicted dynamics. Here, we discuss three primary sources of error: finite data, imperfect data, and incomplete basis functions. The resulting errors can be minimized by adapting steps 2.2 and 2.3 of the scheme above. Additional sources of error can include the recording of unrepresentative DOFs or selecting the wrong equation, which we do not discuss further here.

\textbf{(1) Finite data:} realistic data sets consist of a finite number of trajectories of finite length, with a total length of all trajectories that we call $\tau$. The presence of noise and the potentially only partially explored phase space leads to sampling errors, which are random errors, and therefore vanish for $\tau \to \infty$. 

\textbf{(2) Incomplete basis functions:} In step 2.2 of the inference procedure, a set of $N_b$ basis functions $\{ c_\alpha(x,v) \}$ needs to be chosen to perform the inference. If the basis is not sufficient to accurately represent the underlying model, then even with perfect, infinite data, there will be a systematic error, i.e. a representation error. 

To deal with problems 1 and 2, there is a basic trade-off: as the number of parameters of the basis $N_b$ increases, the representation error decreases, but the sampling error increases. Indeed, the mean-square error (MSE) due to finite data in the estimate of the deterministic term grows linearly with the number of parameters $N_b$, for both under- and overdamped dynamics~\cite{Frishman2018,Brueckner2020}:
	\begin{equation}
	\label{error}
	\mathrm{MSE} \propto N_b/ \tau
	\end{equation}
On the other hand, the representation error decreases with $N_b$ in a way that depends on the underlying model and the set of basis functions considered. Thus, for a given amount of data $\tau$, there is an optimal basis size $N_b^*(\tau)$ that can be inferred. To connect this discussion to the inference approach for confined migrating cells described in section~\ref{sec:two_state}, we point out that the grid-based binning approach effectively corresponds to a basis of top-hat functions at regularly spaced locations in the phase-space (Fig.~\ref{fig:two_state_inference}b). The parameters $F_{\alpha}$ then correspond to the average acceleration $\dot{v}$ at that location in phase-space. This approach requires a large number of fitting parameters $N_b$ and would therefore not perform well in high-dimensional systems such as interacting cells or experimental data sets with low statistics.  In the confined cell example, we had access to a large data set of 1D nucleus trajectories (Fig.~\ref{fig:two_state_inference}A), and thus the binning approach was feasible in this case. A generally better approach is to use a set of \emph{smooth} basis functions, such as polynomials or Fourier components. In this case, fewer parameters are required and additional constraints such as symmetries of the system can be taken into account to further restrict the choice of basis functions. We will discuss how this approach allowed inferring an equation of motion for the more complex case of interacting cells in section~\ref{sec:towards_collective}. 

Importantly, while Eq.~\eqref{error} provides an argument for how many parameters $F_\alpha$ can be learned from the observed trajectories, it does not tell us which parameters are relevant to describe the dynamics. For this, sparsity-enforcing methods, such as such as SINDy~\cite{Brunton2015,Champion2019}, are required. Briefly, these work by augmenting the cost function that is minimized by the inference by penalizing non-zero parameters. Such sparsity constraints have recently been generalized to stochastic systems~\cite{Boninsegna2018,Dai2020a,Callaham2021,Huang2022}, but have not yet been applied to experimental data sets.

\textbf{(3) Imperfect data:} experimental trajectories are inevitably subject to measurement error, and are recorded at finite time intervals $\Delta t$. Both of these in principle separate effects -- discreteness and measurement error -- lead to systematic biases in the inference results, which rely on the numerical derivatives of the trajectories. To address these issues, estimators with bias-corrections have been developed.

To illustrate this, we first focus on how to deal with discreteness in an underdamped system~\cite{Brueckner2020a}. This poses a non-trivial problem, since if only the trajectory $x(t)$ is observed at discrete time steps, this means that one of the dynamical variables, the velocity, is not observed, but has to be estimated as well. Importantly, this leads to systematic errors to the inference result, which persist even in the limit as $\Delta t \to 0$, and do not average away even in the limit of infinite amounts of data. For instance, in the simplest case of an underdamped equation of motion with a linear damping term, i.e. the persistent random motion model (Eq.~\eqref{OU_eom}), the conditional average of the accelerations does not converge to $-v/\tau$, but to $-\frac{2}{3}v/\tau$, as first pointed out in ref.~\cite{Pedersen2016}. This is why we wrote Eq.~\eqref{OU_condav} as an approximation rather than an equality. Thus, this systematic error has to be addressed by de-biasing the estimator of the deterministic term. This estimator can be derived by inverting the stochastic Ito-Taylor expansion of the equation of motion~\cite{Kloeden1992}. The estimator for the deterministic term that is robust against discretization effects then reads~\cite{Brueckner2020}
	\begin{equation}
	\label{estimator_F}
	F_{\alpha} = \langle \dot{v} c_\alpha(x,v) \rangle - \frac{1}{6} \left\langle \sigma^2 \partial_v c_\alpha(x,v) \right\rangle
	\end{equation}
Here, the first term is the conditional average of the accelerations (equivalent to Eq.~\eqref{OU_eom}), while the second term is a correction term that arises due to the projection of an estimated second derivative (acceleration) onto an estimated first derivative (velocity) of the degree of freedom. Note that the correction term vanishes for deterministic systems ($\sigma = 0$) and depends on the derivative of the basis function, indicating that smooth basis functions should be used. This correction can therefore not be applied to a binning inference. For the stochastic noise amplitude the estimator
        \begin{equation}
	\label{estimator_sigma}
	\sigma_{\alpha}^2 = \frac{3 \Delta t}{2} \left\langle \dot{v}^2 c_\alpha(x,v) \right\rangle 
	\end{equation}
which depends on the square of the accelerations, akin to a standard deviation of the accelerations, since at short time-scales, the accelerations are dominated by the noise. Another way to address the error due to discretization is a Bayesian maximum likelihood approach~\cite{Ferretti2020}.

A conceptually similar approach can be used to de-bias estimators from systematic errors due to measurement errors. In this case, the assumption is that measurement errors are uncorrelated in time, meaning that multi-point estimates of the numerical derivatives can be combined in such a way that all systematic error terms vanish by construction~\cite{Frishman2018,Brueckner2020}.

In summary, rigorous stochastic inference approaches allow inference of the governing dynamical systems from observed experimental trajectories in a way that trades-off finite data limitations and representation complexity, and is robust to measurement errors, and discretization errors. That said, there are many open challenges in terms of method development that could aid future applications of stochastic inference to cell migration data. For instance, there is currently no principled approach to automatically identify the most representative DOFs directly from an experimental movie (step 1 in section~\ref{sec:inference_principles}), although first approaches towards learning dynamics from movies have been made~\cite{Gnesotto2020}. Furthermore, there is selecting the class of equation to fit is typically done based on physical intuition rather than agnostic approaches (step 2.1 in section~\ref{sec:inference_principles}). Here, generalizing approaches for low-noise systems leveraging Taken's embedding theorem
could be a potential direction. This could also provide a way to address the potential role of unobserved hidden variables in the dynamics, such as the cell polarity. Finally, while approaches for basis selection and sparsity constraints have been applied to stochastic systems~\cite{Boninsegna2018,Dai2020a,Callaham2021,Huang2022}, they have not yet been combined with bias-corrected stochastic estimators, and are therefore not yet applicable to underdamped stochastic systems.

\section{Inferring heterogeneity in cell behaviour}
\label{sec:variability}
A key feature of migrating cells is the large variability of the observed behaviours within a cell population. A typical set of cell migration trajectories exhibits large variations in behaviour between different individual trajectories, but also over time within a single trajectory. Tracing the origins of such variability is an open challenge for which data-driven approaches are ideal, since it naturally relies on the analysis of large ensembles of observations~\cite{Vestergaard2015}. Here, we propose to distinguish four distinct contributions to the behavioural variability of migrating cells:
\begin{itemize}
\item \textbf{Intrinsic stochasticity}: the intra-cellular machineries driving cell behaviours operate at the molecular level, and are thus subject to intrinsic noise. At larger scales, where these molecular degrees of freedom remain unobserved, this intrinsic noise leads to seemingly random patterns in behaviour, which we refer to as the intrinsic variability of cell behaviour.
\item \textbf{Cell-to-cell variability}: even in populations of cells with identical genomes, the stochasticity of intra-cellular processes such as gene expression, cytoskeletal rearrangement and protein localization can lead to large differences in the proteomes of individual cells~\cite{Niepel2009,Sigal2006,Raj2008,Wagner2016,Altschuler2010}. At the cellular scale, this diversity can lead to variations in cell behaviour, which is also referred to as phenotypic or population heterogeneity. The connection from molecular to behavioural heterogeneity has been demonstrated in cellular processes ranging from growth rate and drug response to morphology~\cite{Cohen2008,Feinerman2008,Gascoigne2008,Wieser2009}, and has been suggested to play an important role in collective cell migration~\cite{Camley2017,Li2019}.
\item \textbf{Temporal variability}: the behaviour of cells may also exhibit variations over time: as cells undergo the cell cycle, they grow, which may also affect other behaviours, including cell migration~\cite{Boehm2001}. Furthermore, cells may switch between qualitatively distinct modes of behaviour, meaning that separate models for each behaviour, as well as a model for the switching itself, must be considered.
\item \textbf{Environmental variability}: potentially unobserved changes in the extra-cellular environment may cause changes in behaviour, which could be mistaken for other types of variability. This can occur, for example, in cell migration experiments in environments with unobserved structures, including porous 3D matrices~\cite{Wu2014}.
\end{itemize}
Clearly, these different sources of variability are hard to disentangle, and sometimes it may not be clear, even in principle, what should be counted as intrinsic noise vs temporal variability. Here, we argue that the key property is the time-scale of deviations of the process from the ensemble- and time-average. The definition of intrinsic noise is that it is stationary in time, implying no long-lasting deviations from the ensemble and time average. Cell-to-cell variability, in contrast, by definition leads to deviations on a time-scale equal to the lifetime of a single cell: these are differences in behaviour of a single cell relative to the population average that persist for the entire life of the cell (and may in principle be passed on to daughter cells). Temporal variability on the other hand leads to deviations on time-scales similar to the time-scales of the behaviour of interest, but shorter than the lifetime of the cell. 

The variability of cell behaviour can make model inference challenging. Firstly, significant variability can mean that a large data set of trajectories is required to sufficiently sample the distribution of behaviours to infer a reasonable ensemble- and time-averaged model. Secondly, in some cases, ensemble- and time-averaged models may not be predictive of behavioural statistics that are sensitive to the variability. In this section, we will focus on inference approaches that are specifically tailored to quantify and characterize the variability of observed cellular behaviours in single-cell contexts. Gaining insights into how these distinct contributions determine the overall variability of cell migration could be important for understanding physiological migration processes, as well as the mechanistic basis of the behaviour. However, disentangling these different contributions to the behavioural variability based on an observed data set can pose a formidable challenge. To this end, several data-driven approaches have been developed in the context of cell migration, which we will discuss here. 

\begin{figure}[h!]
	\includegraphics[width=\textwidth]{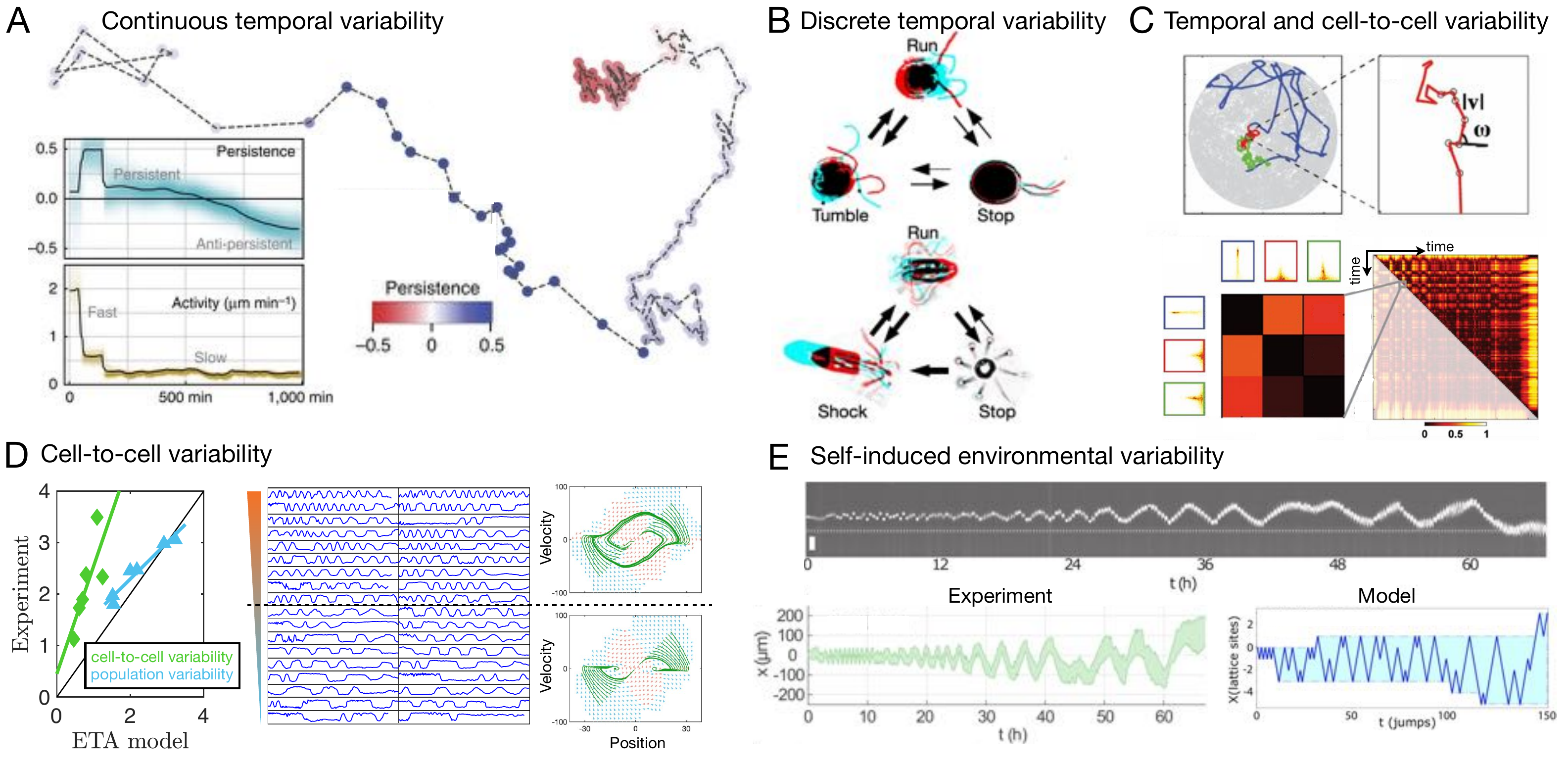}
	\centering
		\caption{
		\textbf{Experimental and theoretical approaches for detecting variability in cell behaviour.}
		\textbf{(A)} For cell behaviour with a variability that evolves continuously in time, the super-statistical random walk analysis infers the persistence and activity of the cell as a function of time. Inset shows the inferred time-dependence for the trajectory on the right (MDA-MB-231 breast cancer cell migrating on uncoated plastic). Adapted from~\cite{Metzner2015} (CC-BY).
            \textbf{(B)} Single-cell ciliate organisms often switch between discrete states of motion, as shown here for \textit{Chlamydomonas reinhardtii} (top, 2 cilia) and \textit{Pyramimonas octopus} (bottom, 8 cilia). Adapted from~\cite{Bentley2022} (CC-BY).
            \textbf{(C)} Trajectories of \textit{Tetrahymena} cells (hundreds of cilia), that are quantified by speed $|v|$ and angular velocity $\omega$ (top). Changeability matrix quantifying the similarity of the joint distribution $P(|v|,\omega)$ over time and across individuals. Adapted from~\cite{Jordan2013}.
            \textbf{(D)} Comparison of experimental statistics to ensemble- and time-averaged (ETA) model predictions reveals the cell-to-cell variability in cell hopping behaviours in two-state micropatterns (left). Green data points: variance of average hopping times across cells; blue data points: variance across the population. Sorting cells by hopping activity reveals distinct underlying deterministic motility patterns (right; MDA-MB-231 breast cancer cells).
            \textbf{(E)} Self-induced variability through protein deposition of a migrating cell exhibiting oscillations of increasing amplitude (top and left; MDCK cells on fibronectin coated 1D lines). A model of the cell as a persistent self-attracting random walk captures the behaviour (right). Adapted with permission from~\cite{DAlessandro2021} (CC-BY).
				 }
	\label{fig:variability}
\end{figure}

\subsection{Quantifying temporal and cell-to-cell variability in behaviour}
\label{sec:variability_behaviour}

Models for cell migration are typically formulated as stochastic equations of motion, which is a natural way to capture processes exhibiting fluctuations. In the equations of cell motion introduced in previous sections (Eqs.~\eqref{OU_eom} and~\eqref{Fxv_eom}), the stochastic white noise term ensures that no two trajectories look alike. This is a model for the \emph{intrinsic stochasticity} of the migration process. To determine the structure and parameters of cell migration models, the dynamics are typically averaged across different cells and over time, yielding ensemble- and time-averaged stochastic models that describe the average member of a cell population. Therefore, these approaches fail to capture cell-to-cell and temporal variability. Similarly, bottom-up models for cell motility typically assume that all cells in a population can be described by a common set of parameters that are constant in time. To demonstrate how data-driven methods can help quantify temporal and cell-to-cell variability, we will first discuss two examples of quantifying temporal variability, before discussing another approach to disentangle the contributions of temporal and cell-to-cell variability.

To develop a framework which can account for temporal variability in cell migration, Metzner et al.~\cite{Metzner2015} developed a generalization of the persistent random motion framework, which allows for time-dependent migration parameters (Fig.~\ref{fig:variability}A). In this `super-statistical' approach, both the persistence $\tau_\mathrm{p}$ and the noise amplitude $\sigma$ in Eq.~\eqref{OU_eom} become functions of time. The values of the parameters are inferred from experimental trajectories using a Bayesian maximum likelihood approach. With this method, the local persistence and activity of migrating cells could be identified as a function of time, revealing pronounced phases of `run' and `rest' states in trajectories of individual cancer cells (Fig.~\ref{fig:variability}A). Such switching between behaviours has been suggested previously to be due to distinct transient intra-cellular organizations~\cite{Maiuri2015}. Interestingly, similar discrete switching between persistent and anti-persistent motion was recently identified \textit{in vivo} in \textit{Drosophila} hemocyte migration using a data-driven machine learning approach~\cite{Korabel2022}. In the future, connecting such data-driven identification of temporal variability to live imaging of intra-cellular features could provide a way to link cellular behaviour to the underlying mechanisms, and how these control switching between subclasses of motility behaviours.

Links between morphological features and behaviour are clearer in the behaviour of swimming protozoans, whose flagella can organize into distinct states corresponding to behaviours such as run, tumble and stopping~\cite{Sasso2018,Wan2018} (Fig.~\ref{fig:variability}B). These distinct morphological states already suggest the existence of discrete behaviours, which is in contrast to the more continuous state space of eukaryotic cells which undergo much more complex shape changes. This discreteness already suggests temporal variability with switching between different modes of behaviour as a natural framework for protozoan motility. Based on trajectory data of such swimming protozoans, quantitative frameworks for migration variability have been developed that characterize the time-dependent motion in sliding time windows to access the joint probability distribution of speed and turning moment~\cite{Jordan2013} (Fig.~\ref{fig:variability}C). By quantifying the change in this distribution over time and between individuals gives rise to a changeability matrix allowing pairwise comparison between any two time points and between individuals. Using clustering and dimensional reduction, this matrix leads to a low-dimensional behaviour space revealing two-state `roaming and dwelling' model of swimming behaviour for multi-ciliate \textit{Tetrahymena} cells. Conceptually similar approaches have been developed for other organisms and their interactions with confining boundaries~\cite{Bentley2022}, revealing different types of discrete cell states including run-tumble-stop behaviour in biflaggelate vs run-shock-stop behaviours in octoflagellate protozoans (Fig.~\ref{fig:variability}B). These analysis frameworks could have potential also for eukaryotic cell migration, provided that a sufficient time-resolution can be achieved experimentally, which is key for a sufficient sampling of the sliding time windows of such an approach. Indeed, we tested the changeability approach~\cite{Jordan2013} on confined cell migration data of cancer cells~\cite{Brueckner2019}, but found that the frame rate was insufficient to properly sample the changeability matrix. Conceptually similar frameworks have also been invoked in the literature on animal behaviour, including fitting of locally linear dynamical systems to motility data of the nematode \textit{C. elegans}~\cite{Costa2019}. Since the data analysis problems in animal behaviour are often very similar to those in cell migration, potentially connecting these approaches to cellular data could be an interesting perspective. In summary, these works provide computational tools to rigorously identify and characterize cell-to-cell and temporal variability in migration behaviours from trajectories alone.

While these frameworks provide a way to quantify and characterize cell migration variability over time and between individuals, they do not provide a method to determine whether such variability exists in the first place. Indeed, if the observed trajectories are short, as is often the case in cell migration experiments, they may appear variable simply due to the randomness introduced by intrinsic stochasticity. How then can real variability be distinguished from apparent variability due to the intrinsic stochasticity? This question has previously been raised in the context of collective cell migration~\cite{Schumacher2017}, where it was suggested to compare the observed variability to an appropriate `null-model'. Specifically, this means performing a direct comparison of variability-sensitive experimental observables, such as population variances, to the predictions by a parameter-optimized model without variability. Deviations from the variability-free model can then provide an indicator for real variability. A difficulty in applying this approach is that it requires both a large ensemble of migration trajectories in a standardized setting, and an appropriate theoretical framework to provide a null-model. 

To demonstrate how such a null-model approach can work in practice, we will show how to use the inferred equation of motion for confined cell migration~\cite{Brueckner2019} (section~\ref{sec:two_state}) as a benchmark to identify behavioural variability~\cite{Brueckner2020}. The inference of this equation of motion was based on the assumption that there is no variability between cells or in time, such that we used an ensemble- and time-averaged (ETA) inference approach. This ETA equation of motion model correctly captures the ETA statistics of the experiment, such as correlation functions~\cite{Brueckner2020}. Thus, this equation of motion provides a null-model to predict the amount of variability between individual (short) trajectories based on only intrinsic noise, to which we can compare the experiment. We found that the variance in behaviour between individual cells was larger in the experiment than that measured in an ensemble of trajectories of similar length predicted by the null-model (Fig.~\ref{fig:variability}D, left). This indicated that there is real cell-to-cell variability in the system, beyond the random variations expected from a single, ergodic and stationary process. Interestingly, our analysis further revealed that within the cell population, there are qualitative differences in the class of dynamical systems describing the migration of individual cells (Fig.~\ref{fig:variability}D, right). Faster cells exhibited limit-cycle dynamics, while slower cells exhibited bistability, with two stable fixed points. The coexistence of distinct dynamical system states within a population of migrating cells has been suggested to originate from a heterogeneity in microscopic migration parameters~\cite{Ron2020a}. Specifically, it was suggested that tuning the elasticity and adhesiveness of cells could lead to distinct dynamical behaviours, including smooth migration, stick-slip migration, as well as bistability between these two modes.

Taken together, these results demonstrate that combining systematic inference tools that account for cell-to-cell variability with mechanistic models could in the future lead to novel insights into the behavioural variability of cell populations. An exciting approach in this respect would be to correlate variability at the molecular scale with variability at the behavioural scale, which could give insight into how molecular organization correlates with behaviour without relying on artificial perturbations of the system~\cite{Chan2017,Todorov2019,Li2013,Sachs2005}.

\subsection{Identifying sources of environmental heterogeneity}
\label{sec:variability_extrinsic}

Conceptually, we think of temporal and cell-to-cell variability to have their underlying cause in cell-intrinsic properties that change over time or between cells, such as protein concentrations or localization. However, in addition to this, migrating cells also encounter variability in their environment including heterogeneous extra-cellular matrices~\cite{Beroz2016,Berthier2022} or contact with other cells~\cite{Carmona-Fontaine2008,Stramer2017}. Indeed, apparent cell-to-cell variability in collective systems has in many cases been shown to be caused by environmental factors, including local cell density, cell-cell contacts and relative location in a cell cluster~\cite{Colman-Lerner2005,Cohen2008,Snijder2009,Snijder2011}. In the context of single cell migration, this was nicely demonstrated by applying the `super-statistical' approach introduced in the previous section to cells migrating through series of constrictions, showing how cell activity and persistence adapt to the local structure of an external confinement~\cite{Metzner2015}. If these external features were not observed, then these responses would appear as strong temporal variability, even though cellular responses to confinements can be explained through an ensemble averaged model that takes into account the cell position within the confinement (as shown in section~\ref{sec:two_state} and ref.~\cite{Brueckner2019}). An interesting special case of such extrinsic variability is \emph{self-induced environmental heterogeneity}, where the cell itself causes changes to its environment which in turn affect its behaviour. These are by nature harder to observe experimentally, and can therefore be mistaken for temporal variability. Here, we discuss two examples of this case in which quantitative frameworks for such self-induced environmental changes were developed. 

First, in 3D migration through a matrix, some cells perform proteolysis, which is a mechanism that allows cells to locally digest the surrounding matrix to create a migration path. This behaviour was shown to lead to asymmetries in the preferred direction of motion of cells: cells were more likely to turn around by $180^\circ$ than expected based on persistent random motion, thus backtracking on their previous path~\cite{Wu2014}. As a model for this process, the \emph{anisotropic persistent random walk} model was proposed, which includes spatially anisotropic parameters (parallel vs orthogonal to the direction of motion) and thereby account for this effect. Such proteolytic behaviour was also shown to lead to directional random walks in the presence of global strain applied to the matrix~\cite{Dietrich2018}.

In the second example, it was found that migrating cells deposit material on the surface on which they migrate, causing them to behave differently when they return to a location that they previously visited~\cite{dAlessandro2021} (Fig.~\ref{fig:variability}E). Specifically, cells were observed to preferentially occupy previously visited areas. In this work, data-driven inference was used to generalize the phase-space analysis introduced in section~\ref{sec:two_state} to the problem of self-attracting migration on a 1D line. This approach revealed that cells deterministically accelerate away from the boundaries of previously explored space. This observation motivated a quantitative description using a \emph{persistent self-attracting walk} model, which quantifies the relative probabilities of turning back vs. exploring new areas. This effect leads to long-lived spatial memory in the migration, which can have dramatic consequences for the ways in which cells search and explore space. Indeed, a phase-field model approach modelling the interaction of cells with their secreted footprint predicted that such memory lets cells switch between confined, oscillatory, and exploratory migration when they explore 2D spaces~\cite{Ipina2023}.

In summary, these approaches identified important cell migration mechanisms using data-driven analysis of the migration trajectories which exhibited striking variability. The analysis revealed that the observed variability is in fact due to extrinsic effects, albeit regulated by the cell itself. These findings are particularly interesting in the broader picture of regulated cell-to-cell variability proposed in~\cite{Snijder2011}, where it was suggested that deterministic, regulated variability could have functional importance in cell population, which is in contrast to cell-to-cell variability caused by random fluctuations of intra-cellular processes.

\section{Connecting cell dynamics to mechanisms}
\label{sec:mechanism}

In the previous sections, we have discussed how quantitative frameworks for cell migration can provide data analysis tools and yield conceptual frameworks to think about cell behaviour. A third important contribution such frameworks could make to the field is by providing constraints for mechanistic cell migration models. We refer to models as `mechanistic' if they are based on a bottom-up approach in which the model is postulated based on known cellular processes and their simplified physical description. This is in contrast to the data-driven, top-down approaches that we have focussed on so far in this review. 

There is a long history of mechanistic biophysical modelling of cell migration (see e.g. ref.~\cite{Danuser2013} for a review). Here, we focus on how combining bottom-up models with top-down data-driven approaches can help address some of the key challenges in the field: (1) Constraining mechanistic models that make predictions for the long time-scale behavioural dynamics of cells. (2) Understanding how cell dynamics may respond to external inputs, and how this could be included in physical models. (3) Connecting different classes of mechanistic models across scales into a coherent theoretical framework for cell migration. We first provide a brief overview the key types of mechanistic models for single cell migration, and discuss how they may be connected to inference approaches to address these challenges (section~\ref{sec:bottomup_models}). Next, we discuss how performing data-driven inference on more complex cellular features beyond cell trajectories, such as cell shapes and protein localization, could provide a bridge between top-down and mechanistic models (section~\ref{sec:beyond_traj}). We review these approaches by systematically increasing the level of coarse-graining and length scale of the models and observables, and specifically highlight how models and data can be compared at each length scale (Fig.~\ref{fig:models}).

\subsection{Bottom-up models for cell migration}
\label{sec:bottomup_models}

\begin{figure}[h!]
	\includegraphics[width=\textwidth]{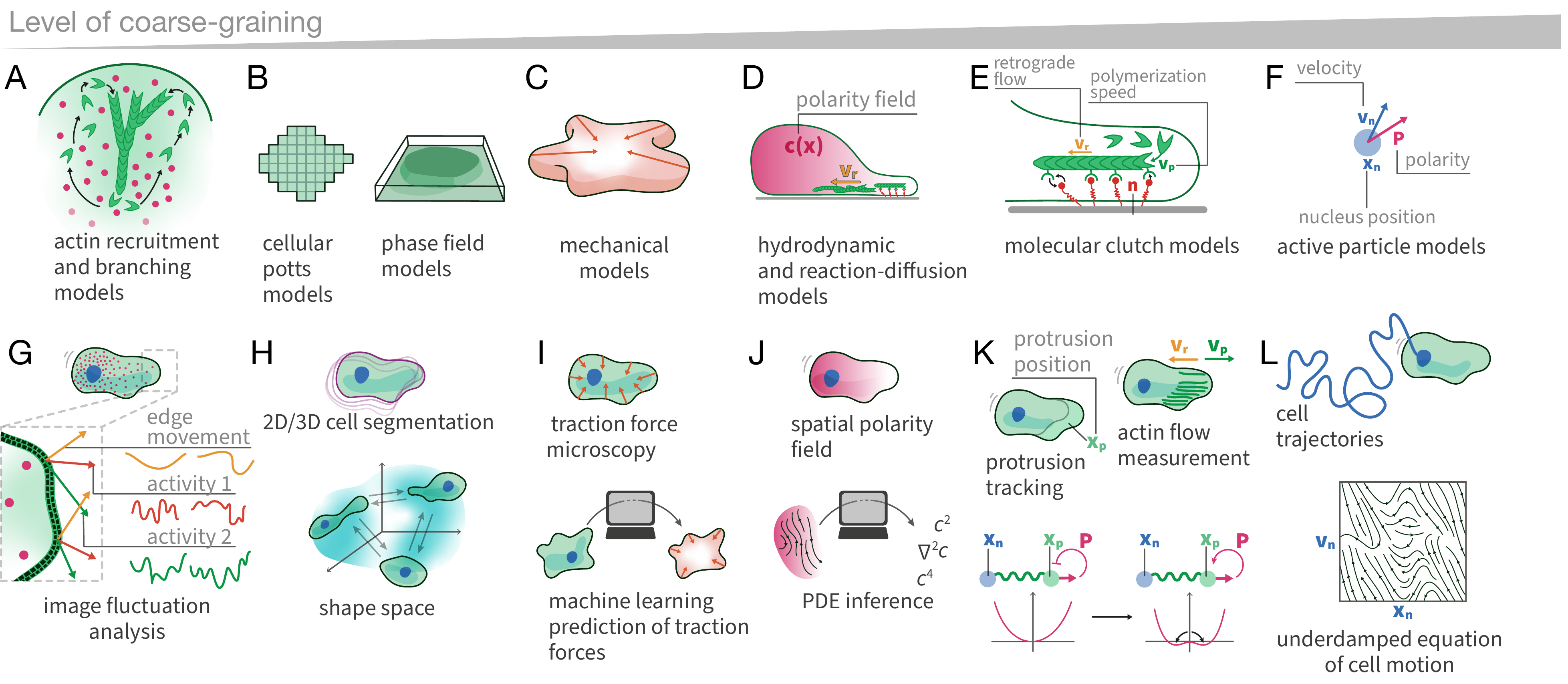}
	\centering
		\caption{
		\textbf{Connecting bottom-up and top-down models for single-cell migration.} \textit{Top row:} bottom-up models for cell migration with decreasing level of coarse-graining from left to right. \textit{Bottom row:} types of cellular datasets and top-down inference approaches at the corresponding levels of coarse-graining.
            (A) Actin recruitment and branching models describe the establishment of the actin network and how it propels the cell membrane in protrusions.
            (B) Cellular potts models and phase field models describe the evolution of the 2D or 3D cell shapes.
            (C) Mechanical models predict cellular traction forces.
            (D) Hydrodynamic models of actin flow and reaction diffusion models describe the coupling of spatially extended polarity fields to motility.
            (E) Molecular clutch models describe how actin retrograde flow and polymerization propel the cell forward by establishing friction with the substrate through adhesions.
            (F) Active particle models describe the evolution of cell position based on velocity and polarity.
            (G) Image fluctuation analysis based on high-resolution live imaging of polarity factors allows causality inference of actin binding factors and how they control protrusion growth.
            (H) Segmentation of the 2D or 3D shape of cells can be integrated in shape space models of morphodynamics, providing a dimensional reduction of complex dynamics.
            (I) Integrating traction force microscopy datasets has been done in machine learning models predicting forces based on protein concentrations.
            (J) Datasets of spatial fields of polarity cues can be treated with partial differential equation (PDE) inference to obtain reaction-diffusion models of cell polarity.
            (K) Protrusion tracking or actin flow measurements allow inference of nucleus-protrusion models.
            (L) Cell trajectory data allows inference of underdamped equations of cell motion.
				 }
	\label{fig:models}
\end{figure}

Bottom-up biophysical modelling of single cell migration initially focussed on particular aspects of the motility machinery, such as the ratchet model for force generation by actin polymerization~\cite{Peskin1993,Mogilner1996}, actin branching~\cite{Mogilner2002}, and the molecular clutch model for adhesion dynamics~\cite{Chan2008} (Fig.~\ref{fig:models}A,E). To integrate these underlying mechanisms into cell-scale models, effective descriptions of their coupling to the large-scale behaviour of cells are required. We will describe these models in order of increasing level of coarse-graining, starting with computational models that account explicitly for the cell shape and subcellular features, and then moving to more coarse-grained approaches describing cells as gels, mechanical modules, and particles (Fig.~\ref{fig:models}).

To couple mechanisms to cell migration at the cell scale and to describe the typical shapes of migrating cells, a number of studies have developed \textit{moving boundary condition models}  (Fig.~\ref{fig:models}A). These models aim to predict the evolution of the cell boundary, thereby predicting both motion and shape. The motion of the boundary can be described by physical models of molecular processes, including polarity signaling~\cite{Maree2006,Satulovsky2008,Nishimura2009}, hydrostatics and membrane tension~\cite{Stephanou2008}, and actin network dynamics~\cite{Shao2010a,Herant2010a}. Zooming out from implementations of cell shape dynamics that assume specific biophysical mechanisms, a popular model providing an effective formulation of cell shape dynamics is the \textit{Cellular Potts Model}~\cite{Graner1992,Glazier1993,Segerer2015,Albert2014,Albert2016b,Albert2016,Albert2016a,Thuroff2019a,Goychuk2018}  (Fig.~\ref{fig:models}B). In this lattice-based model, each cell is described by a set of lattice sites, and the cell shape is evolved by addition and removal of lattice sites based on an energy function which effectively models cell interfacial tension (perimeter elasticity) and a preferred cell area. To implement migration, the energy additionally includes a polarity term. In contrast to this effective energy-based approach, the \textit{Phase Field Model} describes the cell as a field $\Phi(\mathrm{x},t)$ that is equal to 1 inside the cell and 0 outside~\cite{Kockelkoren2003,Ziebert2011,Shao2012a,Marth2014,Camley2014a,Bertrand2020}  (Fig.~\ref{fig:models}B). Cell shape dynamics are then simulated through evolution of this field and coupled to cell polarity and cell velocity through force balance. Other computational cell migration models include the evolution of a cell contour function~\cite{Coburn2013}, models that are based on fiber network implementations of the cytoskeleton~\cite{Zhu2016} and particle-based models with stochastic adhesions to a fibrous extracellular matrix~\cite{Dietrich2018}.

To connect these computational models to experiments, a key challenge is that these models often have many parameters that are difficult to constrain based on experimental data. Furthermore, if a specific mechanism for the cell shape evolution is assumed, it remains unclear how to systematically rule out alternative explanations. Indeed, in a systematic study comparing various computational approaches with different underlying assumptions~\cite{Wolgemuth2011} showed that all these models were able to faithfully capture the migration and typical shape of keratocytes. This apparent degeneracy of possible mechanisms may be a consequence of real redundancy in biological mechanisms responsible for a given behaviour. However, this observation also points to a problem with using complex bottom-up approaches for conceptual insight, as their parameters may remain under-constrained based on phenomenological observations. Therefore, we argue that connecting these computational models to inference approaches can be a promising path to constrain and better understand these models. Specifically, there is no reason why data-driven inference should only be applied to experimental data: we can similarly simulate a computational model, record the trajectories, and learn the effective equation of motion from simulated data. This learned effective model may then provide a much stronger constraint when comparing to the learned model from experimental data. Note that this approach is not limited only to nucleus or center-of-area trajectories of cells, but can equally be applied to more complex cellular features, such as cell shapes, as discussed in the next section. Besides constraining parameters, this connection may allow both better insight into the emergent behaviour in the mechanistic model, as well as the mechanistic basis of a learned behaviour. For example, simulating a Cellular Potts or Phase Field model in confining geometry would allow us to infer how the parameters of these models determine the response of the cell to the confinement; and conversely which mechanistic ingredients are relevant for setting the observed response in the experiment. Thus, connecting these approaches could yield a much more principled approach for constraining mechanistic models and understanding their emergent behaviours.

An alternative route to computational models has been to coarse-grain further and develop simplified, often one-dimensional descriptions of cell polarity and migration dynamics based on underlying physical principles. First, actin polymerization and retrograde flow have been described using continuum theories~\cite{Callan-Jones2008,Hawkins2009,Hawkins2011,Blanch-Mercader2013,Khoromskaia2015,Bergert2015,Lavi2020}, which may be coupled to advection-diffusion models of polarity cue concentrations~\cite{Gracheva2004,Doubrovinski2011,Vanderlei2011,Maiuri2015,Camley2017b} (Fig.~\ref{fig:models}D). Such models have also been extended to account for adhesion-independent cell migration in structured systems where cells actively use friction with the walls or the local topography of the environment to self-propel~\cite{Hawkins2009,Hawkins2010,Reversat2020}. Secondly, the molecular clutch model~\cite{Chan2008} describes the stochastic binding and unbinding of adhesions and their coupling to actin flows (Fig.~\ref{fig:models}E). Simulation models based on the molecular clutch can predict whole-cell trajectories~\cite{Bangasser2017,Prahl2020}. More minimal approaches extended the model to account for the mechano-sensitive binding dynamics of focal adhesions~\cite{Ron2020a,Sens2020,Schreiber2021}, predicting cell behaviours featuring periodic extension-retraction cycles, that have been observed as so-called `stick-slip processes' in the biological literature~\cite{Monzo2016,Hennig2020}. Coarse-graining further, the most minimal models describe cells as active particles with a polarity (Fig.~\ref{fig:models}F). Interestingly, such active particle models can be directly derived from active gel theories, providing a mapping between the two~\cite{Recho2019}.

These one-dimensional cell migration models have given insight into some of the key cellular behaviours observed experimentally. However, it still remains difficult to make predictions for the full stochastic dynamics of cells, and in particular how the intra-cellular mechanisms to structured environments. To connect these models to inference approaches, a promising avenue may be to hierarchically coarse-grain minimal models into a description that is sufficiently simple that it can be inferred directly from experimentally accessible degrees of freedom, such as descriptions of the $(x,v)$-dynamics of the cell. Specifically, an active particle model describing the cell position and polarity as degrees of freedom may be coarse-grained into an equation for the underdamped dynamics of cell velocities, as shown in Eq.~\eqref{eq_overdamped_Fxv}. Thus, the inferred description (e.g. the function $F(x,v)$) can be matched to the dynamics predicted by more interpretable active particle model. This can be challenging as the inferred functions can contain large amounts of features, some of which may be more relevant than others. Furthermore, the functions may not always have a simple analytical form, making the derivation of an exact description difficult. A challenge for future work is therefore to identify ways to link inferred descriptions to bottom-up models in a principled way. Going further, an active gel or molecular clutch model could be mapped into an active particle model, and thereby indirectly linked to the inferable underdamped equation of motion for the cell. For instance, in ref.~\cite{Brueckner2022}, we provided a mapping between a model for the coupled dynamics of cell nucleus, protrusion and polarity and the underdamped equation of motion of the nuclear dynamics alone. Such mappings will be very useful in providing conceptual links between different models, and may help to test and constrain existing models, in particular when they are generalized to non-trivial external confinements. 

A central challenge for bottom-up cell migration modeling is to link the different types of modeling approaches to each other. Ideally, there should be clear mappings between models, allowing to explicitly contrast assumptions and predictions of different approaches. Furthermore, as different models allow descriptions at different levels of detail, there should be a correspondence between the length- and time-scales of the experimental observations or the behaviour of interest, and the type of model employed. Having consistent mappings between models would then allow switching from one model to the other without contradictions. However, the principles of how to link these models together may still be outstanding. To address this, inference methods could help leverage rapidly increasing experimental data sets to constrain how these models fit together. Indeed, beyond inference from cell migration trajectories, expanding the experimentally tracked degrees of freedom, and performing inference on other cellular features such as protrusions, polarities, traction forces or actin flows directly may provide an important tool, which we will discuss in the next section.

\subsection{Inference from cellular features}
\label{sec:beyond_traj}

To gain insight into cell migration mechanisms, a promising emerging avenue is to apply inference approaches to cellular features beyond just cell trajectories, such as imaging of the cell or nucleus shapes, the cytoskeleton or concentrations of focal adhesion and polarity molecules. While we can obtain a self-consistent description of the migration dynamics from the underdamped dynamics inferred from the cell trajectories, this effectively treats all intracellular degrees of freedom as hidden variables. The key hidden variable in the simplest bottom-up model, the active particle model, is the cell polarity. However, cell polarity is notoriously hard to define, and there is no unique, generally agreed-upon molecular marker for cell polarity. In this section, we will discuss various approaches to infer dynamical models from cellular features of increasing detail, starting from cell shapes, to protrusions, polarity markers and traction forces, and point out the recurrent challenge of linking these features to the polarity of the cell.

Analysing the dynamics of cell shapes is attractive as cell shapes are easy to observe experimentally, as simple brightfield microscopy paired with modern segmentation pipelines based on machine learning~\cite{Ronneberger2015,Falk2019} can provide high accuracy, high throughput shape data sets. Clearly, developing a model for the entire cell shape as a function of time is a large jump in complexity from the one-dimensional cell trajectories we have considered so far, and a low-dimensional stochastic equation of motion may not suffice to capture these dynamics. Thus, a first challenge is to determine the dominant contributions to the cell morphology through dimension reduction. By identifying the principle components of the cell shape, recent works have proposed to study cell morphology in a low dimensional space feature space~\cite{Gordonov2016,Tweedy2019,Chan2020,Cavanagh2022} (Fig.~\ref{fig:models}H). From an analysis point of view, these approaches have been very successful by demonstrating that clustering in shape space can be predictive of metastatic potential~\cite{Hermans2013,Elbez2021}, stem cell lineage decisions~\cite{Buggenthin2017} and drug response~\cite{Gordonov2016}, highlighting the rich information content of cell morphologies. Furthermore, morphodynamic feature approaches have allowed comparative mapping of different cell types~\cite{Imoto2021}, identification of migration strategies in 3D matrices~\cite{Eddy2021,Cavanagh2022}, and revealed adaptive switching between different modes of mesenchymal migration~\cite{Shafqat-Abbasi2016}.

However, to build dynamical models of cell morphology, there are two key challenges: (i) to establish a self-consistent shape space for morphodynamics, meaning that future morphological features can be predicted based on the current features, meaning the space is constructed such that the dynamics are Markovian; and (ii) to predict whole-cell motion from the evolution in shape space. Interestingly, in the case of \textit{Dictyostelium} cell morphodynamics, just three principle morphological components on the second timescale were found to be predictive of migration behaviours on the minute timescale~\cite{Tweedy2019}. In contrast, for neutrophil migration, morphological features obtained from different dimension reduction techniques were shown to be insufficient to predict the migration velocity of the cell, suggesting that additional information beyond the shape is required to capture cell polarization in general~\cite{Chan2020}. Thus, to fully resolve cellular dynamics at the level of cell shapes may require the addition of further cellular degrees of freedom which contain polar information.

To include polar information in a morphodynamic feature space, one could include information about the intra-cellular organization, such as the relative position of the cell nucleus or the traction forces, or the velocity of the cell shape. Shape velocity is simply the derivative of the cell shape, corresponding to a `ribbon' of alternating protrusion and retraction areas around the cell. At the scale of individual protrusions, morphodynamic profiling of shape velocities has revealed drastic spatiotemporal heterogeneity at the time-scales of minutes to hours, far below the time-scale of migration~\cite{Machacek2006,Ma2018}, suggesting that high time-resolution experiments are likely to give most insight. To extract the key information of these shape velocities, an alternative approach is to simplify protrusion and retraction areas into a protrusion and retraction `center of mass' of the cell, i.e. a one-dimensional readout (Fig.~\ref{fig:models}K). 

In previous work, we extracted such protrusion trajectories from confined migrating cells, and inferred the coupled dynamics of cell nucleus and protrusion motion~\cite{Brueckner2022}. Interestingly, considering only nucleus and protrusion motion was not predictive of cell motion, and thus a time-correlated polarity-driven protrusion formation was required to capture the dynamics. Constraining the description of these polarity dynamics based on the observed protrusion trajectories revealed that the cell polarity is sensitive to the local geometry of the confinement. Specifically, under strong confinement, the polarity dynamics switches from a negative to a positive, self-reinforcing feedback loop. This geometry adaptation effect leads to a stereotypical cycle of protrusion extension into the constriction, followed by contraction and transmigration of the cell nucleus. The model then predicted, in agreement with experiments, that the protrusion-nucleus cycling disappears when the constriction is removed. This suggests that the positive polarity feedback loop emerges as a consequence of an adaptation of the cellular dynamics to the presence of the thin constriction. By performing inference on data-sets with cellular features beyond the cell nucleus, this approach resulted in equations of motion with mechanistically interpretable terms, including the nucleus adhesiveness, the mechanical nucleus-protrusion coupling, and the coupling of cell polarity to protrusion confinement. Importantly, this model also correctly predicted the inferred underdamped dynamics of the nucleus trajectories alone, providing a link between the more phenomenological approach at the nucleus level to the intracellular polarity dynamics. In further work~\cite{Flommersfeld2023}, the more interpretable protrusion-nucleus model was then used as a prediction target for a mechanistic model of confined cell migration, based on a generalized molecular clutch approach~\cite{Sens2020}. This revealed how membrane tension, actin alignment, and polarity cue diffusion interplay to generate the geometry adaptation effect. 

Beyond cell shape and protrusion dynamics, the shape of the cell nucleus can give important insights into the forces acting during cell migration in 3D confining systems. In cell migration through tight 3D channel confinements (unlike the flat 2D micropatterns discussed in the previous paragraph), the deformation and translocation of the cell nucleus has been shown to be a key rate-limiting step in migration~\cite{Davidson2014a,Denais2016,Fruleux2016,Green2018,Lomakin2019}. To understand how the mechanics of nucleus deformation controls the migration dynamics, measuring the deformation forces acting on the nucleus could yield important insights. However, direct measurement of these forces acting is experimentally challenging. To circumvent this problem, recent work proposed a data-driven approach to infer the deformation force field actin on the nucleus directly from the observed nucleus shapes, relying on a mechanical model of the nucleus as either an elastic solid or an elastic shell~\cite{Estabrook2021}. These inferred forces could then be used to constrain bottom-up models of how the nucleus affects cell migration~\cite{Leong2011,Scianna2013,LeBerre2013,Tozluoglu2013a,Aubry2015,Cao2016}. Taken together, these approaches show how inference from additional cellular features combined with bottom-up mechanistic models can help identify the mechanistic underpinnings of cell migration in complex environments.

A key element that is lacking in inferred models of cellular features is a direct measurement of cell polarity. Potential definitions include the localization of polarity cues such as Cdc42, Rac or Rho GTPases~\cite{Tapon1997} (such as PBD-YFP, a reporter of Rac1/Cdc42 activation~\cite{DAlessandro2021}), the localization of nucleus-actin bindin proteins~\cite{Davidson2020}, or the relative positioning of cell nucleus and organelles such as the Golgi apparatus and the microtubule organizing center~\cite{Nabi1999,Pouthas2008}. These molecular markers are however very challenging to image experimentally at the long time-scales required to obtain coupled migration and polarity trajectories. A more accessible intracellular observable are the traction forces. Recent data-driven work has shown how machine learning can be used to predict these traction forces from intracellular protein localization~\cite{Schmitt2023} (Fig.~\ref{fig:models}I). At an even more molecular level, a common question is often how the different molecular players in protrusion formation and polarity establishment affect each other causally. This is hard to establish based on pharmacological or genetic perturbations, as this usually perturbs the entire network. An alternative approach has been developed using a data-driven method that uses relative temporal correlations of signalling molecule recruitment and actin polymerization within cell protrusions to infer regulatory networks in a perturbation-free manner~\cite{Welf2014,Lee2015a,Isogai2018} (Fig.~\ref{fig:models}G). Extending these ideas to confined systems where cells are monitored on long time-scales, and combining it with stochastic inference methods, could yield key insights into the mechanistic basis of stochastic cell behaviours and their adaptation to the environment. 

\section{Learning the collective dynamics of multicellular systems}
\label{sec:collective}

In physiological contexts, cells do not only interact with their confining extracellular environment, but also with one another~\cite{Poujade2007a,Stramer2005,Weavers2016}. Cell-cell interactions allow cells to organize collective behaviours and thereby address tasks that they could not solve on their own, such as shaping an embryo or healing a wound. Cellular interactions depend on complex molecular mechanisms, including cadherin-dependent pathways and receptor-mediated cell-cell recognition~\cite{Carmona-Fontaine2008,Stramer2017,Astin2010,Davis2015,Moore2013,Matthews2008,Kadir2011}. 
These mechanisms can lead to well-defined, stereotypical cell behaviours upon collision. A prominent type of collision behaviour was discovered in the 1950s by Abercrombie and coworkers~\cite{Abercrombie1954a}, and was termed \emph{Contact Inhibition of Locomotion} (CIL). CIL refers to the tendency of cells to retract their lamellipodia, repolarize, and migrate apart upon contact. While these observations were made in a simple cell culture on 2D substrates, the relevance of CIL for physiological processes was later demonstrated, for example in the development of the neural crest~\cite{Carmona-Fontaine2008,Mayor2010,Stramer2017}.

At larger scales, cell-cell interactions lead to coordinated collective migration, which has been described with a variety of physical modelling approaches. These include active hydrodynamic theories~\cite{Marchetti2013}, vertex~\cite{Honda1983,Fletcher2014,Alt2017}, mechanical~\cite{Serra-Picamal2012}, and mechano-chemical~\cite{Boocock2020} models, cellular automata~\cite{Segerer2015,Ilina2020}, phase-field models~\cite{Camley2014a,Bertrand2020}, as well as active particle models~\cite{Smeets2016,Sepulveda2013,Basan2013,Copenhagen2018,Garcia2015,Dalessandro2017} (see~\cite{Hakim2017,Camley2017c,Alert2020} for reviews). These modelling avenues typically make \emph{a priori} assumptions on the types of interactions between individual cells, and therefore classify as bottom-up approaches. Cell-cell interactions are frequently modelled using repulsive potentials as an implementation of excluded volume interactions, alignment terms~\cite{Sepulveda2013,Basan2013,Copenhagen2018,Garcia2015}, or explicit implementations of CIL-like reorientation events upon collision~\cite{Smeets2016,Dalessandro2017}. However, in these approaches, the structure of these interactions are usually assumed based on physical intuition (i.e. they are bottom-up models), not derived directly from experimental data. Deriving cell-cell interactions directly from data could have several advantages. Bottom-up models can exhibit model degeneracy, meaning that multiple possible mechanistic interactions can reasonably well capture the qualitative cell behaviours. Furthermore, interacting cells can exhibit complex and unexpected types of interactions, which might be missed in bottom-up models that are limited to physics-inspired interactions such as alignment and attraction/repulsion interactions. Therefore, bottom-up models could be complemented by top-down inference of interactions directly from data, providing stronger constraints on such models. 

We want to highlight three key hurdles that make the development of data-driven approaches for cell-cell interactions difficult. First, inference in interacting active many-body systems is technically challenging. To perform inference on such high-dimensional stochastic systems, a number of approaches have been developed that we discuss in section~\ref{sec:learn_coll}. Second, the structure of the interactions between cells may be substantially more complex than typically encountered in active matter theory, such as nonreciprocal interactions or interactions that adapt and change over time. Third, the complexity of the biological settings in which cell-cell interactions take place make it difficult to disentangle the distinct contributions of single-cell behaviour, interaction with the local micro-environment, cell proliferation, and cell-cell interactions. To overcome these problems, studying interacting cells in simplified artificial environments is a promising direction, which we discuss in section~\ref{sec:towards_collective}.

\subsection{Inference approaches for interacting active systems}
\label{sec:learn_coll}

The inference of interactions from experimental tracking data has been a subject of interest in the field of animal behaviour for a long time~\cite{Partridge1981}. The basic problem in inferring collective animal behaviour is very similar to the challenges faced in collective cell migration data, suggesting that these fields could learn from or even help each other. Specifically, in both systems, the basic problem is how to estimate the response of individuals to the presence of another individual as a function of their relative distance and orientation. In the context of the social interactions of fish a number of approaches to address this problem have been developed~\cite{Lukeman2010a,Katz2011,Gautrais2012}. These inference approaches mainly focused on zonal interaction models which infer how the animal response varies as a function of angle at which another animal is observed, due to the key role of the field of vision in animal interactions. 

To learn a predictive model of interacting cellular systems, we require approaches that can infer the interacting equations of motion of the system, which include both single-cell behaviour and interacting terms. A number of such approaches have been developed for deterministic systems~\cite{Lu2019,Miller2020}, as well as for stochastic systems in the context of animal behaviour~\cite{Escobedo2019,Ferretti2020}, and more generally for interacting stochastic active particle systems~\cite{Frishman2018,Brueckner2020}. Performing inference on collective systems is challenging due to the high dimensionality of the problem: a 3D swarm of $N$ particles has $6N$ degrees of freedom (counting only positions and velocities, although more variables may be relevant), and ?curse of dimensionality? arguments make this problem seem intractable. Indeed, the very simple approach of grid-based binning of the phase space (section~\ref{sec:learning_eoms}) is unfeasible as it would require $\gg 6N$ parameters to accurately represent the dynamics, therefore necessitating prohibitively large data sets to constrain such an approach. To overcome this problem, the trick is to adapt the selection of basis functions in such a way that the inference problem becomes effectively low-dimensional and thereby tractable. In the context of a basis expansion (Eq.~\eqref{expansion}), we can think of binning as using top-hat basis functions at regularly spaced locations in the phase-space. To make better choices for interacting systems, essentially only step 2.2 of the inference procedure in section~\ref{sec:inference_principles} has to be adapted, while the other steps are largely unaffected. The key idea is to simplify the inference by assuming symmetries of the interactions that reduce the number of fitting parameters. 

The first important simplifying assumption is to treat particles as identical, such that all particles obey the same equation of motion. With this assumption, systems with more particles actually become effectively easier to infer from, as there is more data per recorded time-step. As proposed in refs.~\cite{Frishman2018,Brueckner2020}, a natural choice of basis functions is then to expand the deterministic contribution to the underdamped dynamics of cell $i$ as a sum of one-body and two-body (interaction) terms:
\begin{equation}
\label{expansion_interacting}
\mathbf{F}_i \approx \sum_{\alpha} \mathbf{F}^{(1)}_\alpha c^{(1)}_\alpha(\mathbf{x}_i,\mathbf{v}_i) + \sum_{\beta} \mathbf{F}^{(2)}_\beta \sum_{i \neq j} c^{(2)}_\beta(\mathbf{x}_i,\mathbf{v}_i,\mathbf{x}_j,\mathbf{v}_j) + ...
\end{equation}
where higher orders can in principle be included to account for multi-body interactions. This provides a generalization of the basis expansion in Eq.~\eqref{expansion} to interacting systems. 

The second important assumption is to choose two-body basis functions $\{ c^{(2)}_\beta \}$, which reflect additional symmetries of the interactions, such as radial symmetry of the interactions. In that case, all position dependence simply becomes a function of the radial distance $r_{ij}$ of each pair of cells. A further promising approach is the inclusion of small convolutional neural networks as the basis functions of the expansion (Eq.~\eqref{expansion_interacting}), which has been applied to the case of interacting active colloids~\cite{Ruiz-Garcia2022}. An advantage of this approach is that it may reduce the risk of overfitting and provide a flexible basis for complex interaction functions. Furthermore, combining symmetry-based constraints with sparse regression~\cite{Brunton2015,Champion2019,Boninsegna2018,Dai2020a,Callaham2021,Huang2022} could provide an avenue to constrain interacting cellular systems, which was recently demonstrated for simulated models of collective cell migration~\cite{Messenger2022}. Taken together, by decomposing the inferred dynamics into single-cell and interaction terms, and constraining these by symmetry provides a solution for the curse of dimensionality problem, which could facilitate inference of cell-cell interactions from experimental collective migration data.

In the context of collective cell migration, a key objective is to identify different classes of interactions that may have biological interpretation. First, it is important to distinguish between two classes of interactions: \textbf{positional (isotropic) interactions}, which only depend on the relative position of two cells, such as excluded volume (repulsion) or adhesion (attraction) interactions. In contrast, \textbf{orientational (anisotropic) interactions} couple the directional properties of cells to each other. These can be either nematic, such as cell elongation, planar cell polarity, or myosin distributions in epithelia, or polar, such as polarity or velocity vectors. Examples for this are polarity and velocity alignment, stress-polarity coupling, and contact inhibition of locomotion~\cite{Alert2020}. Importantly, these different types of interactions imply different couplings of the position, velocity and polarity vectors of the cells. Thus, inferring the structure and parameters of cell-cell interactions with a general ansatz such as Eq.~\eqref{expansion} directly from experimental data could then give insight into which of these interactions are at play.

In the discussion above, cell collectives are approximated as interacting active particles. The validity of this description may depend on the biological context. For low density, non-confluent assemblies of cells with transient interactions (collisions) between cells active particle model have been shown to provide a good approximation to the dynamics~\cite{Dalessandro2017,Zisis2022}. While active particle models have also been successful in describing confluent cell monolayers~\cite{Sepulveda2013}, in this case, the contact geometry between neighbouring cells is important for cell-cell interactions, which is defined by the cell shapes. Thus, accounting for shape is likely more important in the confluent than in the non-confluent regime. Furthermore, cells are consistently connected to each other, implying that interactions may be governed by the topological graph connecting the cells (as a function of neighbour relationships) rather than metric interactions (as a function of distance). Such epithelia are often described using the so-called vertex model, in which cell shapes are represented by a set of vertices that denote the common point of three or more neighbouring cells~\cite{Farhadifar2007,Fletcher2014,Alt2017}. The commonly used energy function of this model contains area and perimeter terms, which depend on the vertices connected to the (on average) six neighbouring cells. A direct inference of the parameters and interactions of such models could yield novel insights into how tissues control the collective states of cells. In general, expanding the overdamped equivalent of Eq.~\eqref{expansion_interacting} could account for these interactions, but it will not provide an adapted basis to the geometry of the problem. Thus, fitting a set of hypothesized dynamical terms (such as the vertex model) could be a better approach. However, cellular states in epithelial tissues are known to undergo dynamical changes, such as changing myosin distributions, and remodeling of the cell edges~\cite{Kasza2007,Wozniak2009}, with key implications for tissue dynamics~\cite{Noll2017,Sknepnek2023,Ioratim-Uba2023}. This implies that the dynamical parameters are non-stationary in time, making the inference problem significantly more difficult.

One avenue to circumvent this problem is to use a basis of modes that uses the geometry of the epithelial layer more directly~\cite{Noll2017,Brauns2023}. This approach applies in a regime that is dominated by tension in the cell edges, and that exhibits a time-scale separation between the tissue dynamics and the relaxation of individual cell edges, implying that the vertices are in force balance. Under these assumptions, the tissue dynamics decomposes into two independent contributions: the tensions determine the dynamics of the angles (at fixed areas), while isogonal modes predict the evolution of areas (at fixed angles)~\cite{Noll2017}. Due to force balance at each vertex, the tensions can then be directly inferred from the experimentally observed angles at each vertex~\cite{Brauns2023}. Thus, in this decomposition, the tensions are no longer (hidden) parameters that have to be fitted indirectly, but can be measured directly from the images. Incorporating simplifying assumptions therefore allowed a decomposition of the dynamics that makes it feasible to reliably infer time-dependent cellular adaptation. We will discuss the implications of such tension inference in more detail in the next section.

Beyond these models of cells as discrete entities, active polar or nematic hydrodynamic models provide important conceptual frameworks to describe cellular assemblies. To learn such models from observed data, inference and machine learning approaches for active nematics have been developed and applied in the context of \textit{in vitro} microtubule assays~\cite{Colen2021,Golden2023}, and active polar particle experiments~\cite{Supekar2021}. These approaches use the observed velocity fields or cell tracking data to uncover the hydrodynamic equations governing these active matter systems, which could provide a promising approach for inference from collective cellular systems.

In summary, the development of these inference approaches that can be applied to stochastic interacting active systems opens up new avenues to learn the dynamics of cell-cell interactions directly from observed data, and we will review such applications in the next section.

\subsection{From cell pairs to collective migration}
\label{sec:towards_collective}

\begin{figure}[h!]
	\includegraphics[width=\textwidth]{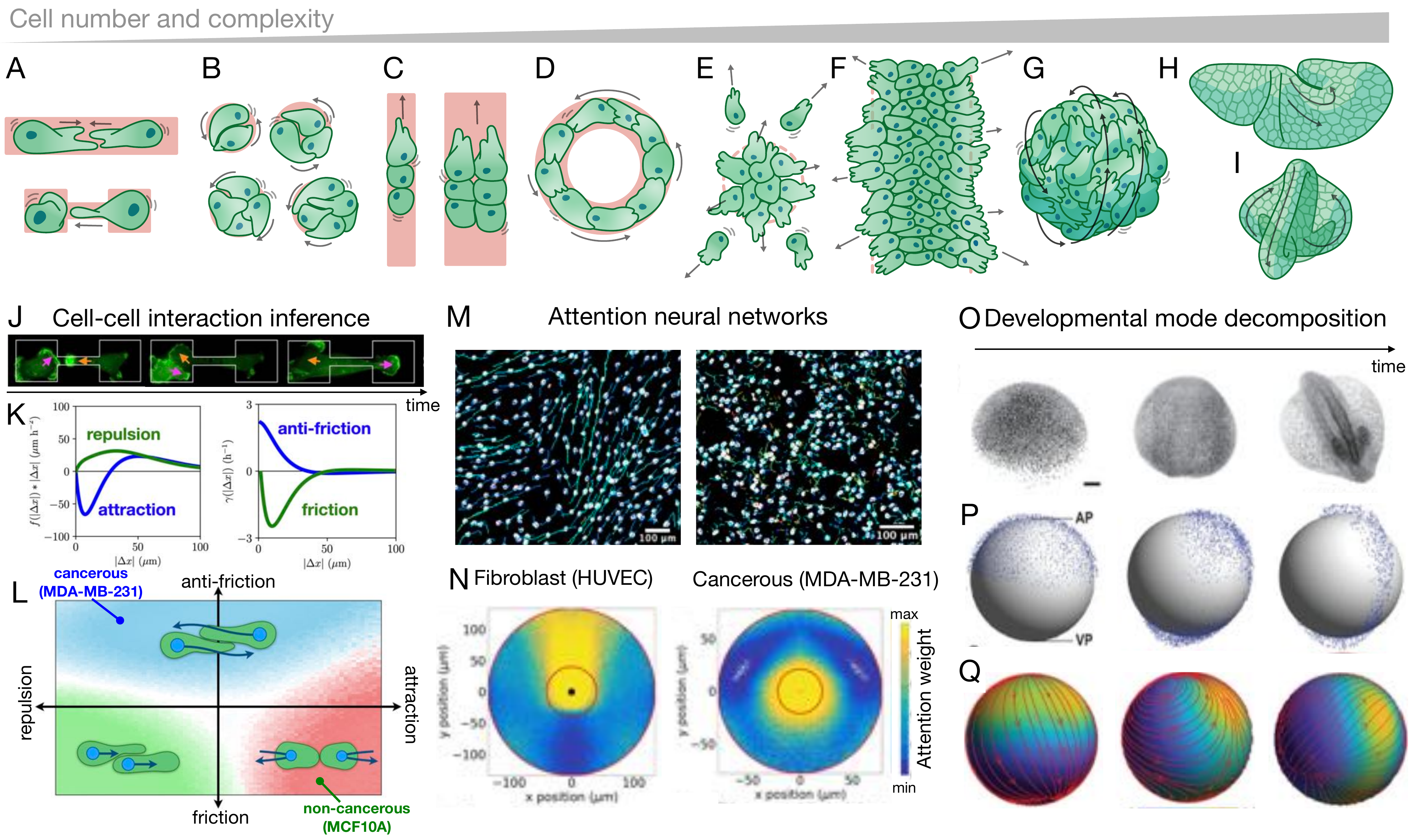}
	\centering
		\caption{
		\textbf{Learning cell-cell interactions from collective cell migration experiments.}
		(A-I) Systems to study collective cell migration with increasing number of cells and complexity: cell pairs (A); circular micropatterns (B); linear (C) and periodic (D) microstripes; collective escape from a confined area (E); spreading monolayers (F), 3D spheroids (G) and embryos such as fly (H) and zebrafish (I).
            (J) Time series of two interacting MDA-MB-231 cells transfected with LifeAct-GFP. Arrows highlight regions of pronounced actin activity, and the arrow color indicates the cell identity.
            (K) Inferred cohesive interaction term $f(r)\Delta x$ (left), and effective frictional interaction term $\gamma(r)$ for non-cancerous MCF10A cells (green) and cancerous MDA-MB-231 cells (blue). 
            (L) Interaction behaviour space predicted by varying the amplitude of the cohesive and friction interactions. Color and cartoons indicate the dominant collision behaviour at each point, and the inferred locations of both cell types are indicated. Adapted from~\cite{Brueckner2021}.
            (M,N) Representative cell trajectories within living tissues, from human umbilical vein endothelial cells (HUVEC, left) and metastatic human breast cancer cells (MDA-MB-231, right). Scale bars: 100 $\mu$m. Normalized attention weight contour plots for both cell types. Red circle: radius of the closest neighbor location. Adapted from~\cite{LaChance2022} (CC-BY).
            (O-P) Mode decomposition of early zebrafish development. Microscopy images (O, scale bar: 100 $\mu$m) and tracked cell positions (P, blue dots) as a function of developmental time. (P) Density (color) and flow fields (arrows) corresponding to the same images. Adapted from~\cite{Romeo2021} (CC-BY).
				 }
	\label{fig:collective}
\end{figure}

Having discussed approaches to make inference from active interacting systems tractable, we turn to the second challenge: reducing the biological complexity of interacting cellular systems, to make the decomposition into various contributions from single-cell dynamics, proliferation, and interactions tractable. To achieve this, a broad variety of \textit{in vitro} approaches to confine groups of cells to defined geometries have been developed. In some cases, this confinement is kept for the entire duration of the experiment, while in others it serves as an initial condition from which cell spreading is observed. To highlight how cell-cell interactions can be inferred from these different approaches, we will systematically go up in the complexity of the experimental approaches, such as increasing cell number and dimensionality of the system, and highlight how interaction inference was or could be applied to these systems (Fig.~\ref{fig:collective}A-I). Note that our discussion here is primarily focussed on substrate-dependent migration, where cells are exerting active migration forces on a substrate in the direction of their polarity. We will briefly touch on the case of substrate-independent tissue flow due to active stresses exerted between the cells through their junctional actomyosin (rather than between cells and substrate via protrusions) at the end of this section.

Many systems that are controlled by cell-cell interactions rely on the concerted dynamics of small groups of cells, including the dynamics of pairs of cells~\cite{Carmona-Fontaine2008,Kozak2023} and migration of small clusters~\cite{Bianco2007,Dai2020} in developmental systems, as well as migrating tumour clusters of up to eight cells~\cite{Friedl1995,Hou2011}. Studying cell-cell interactions at fixed total number of cells $N$ in simplified \textit{in vitro} systems provides a major simplification, as the dynamics of cell proliferation can be neglected, and allows building complexity step-by-step.

The simplest possible cell-cell interaction system is to keep $N=2$ and study the collisions of pairs of cells. This has been done by studying cell collisions on 1D micropatterned tracks~\cite{Milano2016,Li2018,Desai2013,Scarpa2013,Mohammed2019a}, microfluidics~\cite{Lin2015}, and suspended fibers~\cite{Singh2020}. Furthermore, cell pairs have been confined into closed confinements in which they continuously interact over long periods of times, including circular~\cite{Huang2005} and two-state micropatterns~\cite{Brueckner2021} (Fig.~\ref{fig:collective}A,B). An advantage of such closed confinements is that cells interact with each other repeatedly in a standard environment, leading to long interacting trajectories.

In previous work, we applied stochastic inference to such interacting cell pair trajectories to learn the positional and orientational interactions of the cells. To separate these interactions, we used a simplified version of Eq.~\eqref{expansion_interacting} and postulated that the dynamics of the system can be described by the following equation of motion:
\begin{equation}
\label{2cell_eom}
\frac{\d v}{\d t}  = F(x,v)+ f(r)\Delta x + \gamma(r)\Delta v + \sigma \eta(t)
\end{equation}
where $r=|\Delta x|$ is the distance between the two cells and we simplified the dynamics to one dimension since cells move predominantly along the $x$-direction of the pattern. This approach assumes that the deterministic dynamics of the system can be decomposed into two separate components: a single-cell term $F(x,v)$, similar to that inferred from single-cell experiments, and interactive components, which depend on the relative position $\Delta x$ and the relative velocity $\Delta v$ of the cells. The term $f(r)$ thus represents positional cell-cell interactions such as repulsion and attraction. In contrast, $\gamma(r)\Delta v$ depends on the relative motion of the cells, and is therefore an orientational interactions with the mathematical form of an effective frictional interaction. For $\gamma <0$, this interaction accounts for alignment between cells, as it seeks to minimize differences in relative velocity~\cite{Sepulveda2013,Basan2013}. To infer this model (Eq.~\eqref{2cell_eom}), we use a basis expansion of the two-body terms $\{ c^{(2)}_\beta \}$ that assumes radial symmetry of the interactions as the kernels $f$ and $\gamma$ only depend on the distance $r$. The basic inference procedure for this case is the same as for single cells: we infer the model using a suitable basis expansion, and then make predictions for long time-scale statistics of the interacting dynamics, which match those observed experimentally. Importantly, we also find that the single-cell term inferred from two-cell experiments matches that inferred from single cell experiments, suggesting that the separation of interactions and single-cell behaviour was successful.

Interestingly, the inference revealed that a non-cancerous (MCF10A) and a cancerous (MDA-MB-231) breast tissue cell line exhibit distinct types of interactions: while the MCF10A cells exhibit repulsive and regular frictional interactions, the MDA-MB-231 attract at short distances and exhibit a positive friction term ($\gamma >0$) (Fig.~\ref{fig:collective}K). This `anti-friction' interaction ensures that rather than slowing down upon collision, cells deterministically accelerate, leading to the characteristic sliding events observed for this cell line. The model (Eq.~\eqref{2cell_eom}) furthermore suggests  an `interaction behaviour space', which relates the physical interaction terms to the cell-cell collision behaviour of the system, suggesting that this framework could potentially describe various cell-cell interaction modes known in the biological literature~\cite{Stramer2017,Abercrombie1979,Milano2016,Li2018,Hayakawa2020}, including reversing, sliding and following interactions (Fig.~\ref{fig:collective}L). These inferred interactions could in the future provide  contraints for bottom-up models of cell pair collisions~\cite{Singh2020,Zadeh2022}

Going beyond cell pair dynamics, several studies have systematically investigated the effects of increasing the number of interacting cells one by one. First, on circular micropatterns going from $N=2$ to $N=8$ revealed cell number-dependent rotational behaviours~\cite{Segerer2015} (Fig.~\ref{fig:collective}B). Secondly, a number of studies have investigated the behaviour of small clusters of cells on linear and circular microstripes~\cite{Vedula2012,Tarle2017,Jain2020,Vercruysse2022,Pages2023,Ron2023} (Fig.~\ref{fig:collective}C,D). For instance, in trains of keratocyte cells confined to one-dimensional stripes, the speed of the train was shown to be independent of the number of cells in the axial direction, i.e. parallel to the direction of motion~\cite{Vercruysse2022}. In contrast, train speed decreased with cell number in the lateral direction, i.e. orthogonal to the direction of motion. This observation constrained a bottom-up active matter simulation to identify the cell-cell interactions in the clusters. 

The approach of using experimental data to constrain the interactions in bottom-up active particle simulations has been successful in a broad variety of larger scale systems at the scale of hundreds of cells. One strategy to regularize collective behaviour has been to confine cells to a circular confinement which is then released by mechanical or chemical means to allow spreading of cells (Fig.~\ref{fig:collective}E). In \textit{Dictyostelium} colonies, this revealed that a new type of interaction had to be included in an active particle model for the system, which enhances rather than inhibits motility upon collision and was termed Contact Enhancement of Locomotion~\cite{Dalessandro2017}. In cancer cell colonies, a similar experimental and theoretical approach revealed how E-cadherin junctions control excluded volume interactions between cells by `sharpening' inter-cellular boundaries~\cite{Zisis2022}. At the scale of confluent monolayers of cells (Fig.~\ref{fig:collective}F), the cell-cell interactions were captured employing active matter models, using the velocity distributions and correlations as a constraint~\cite{Sepulveda2013}. In these large-scale systems, direct inference of cell-cell interactions is challenging, although recent work proposed a systematic fitting procedure of a Viscek-type alignment model to collective monolayer migration~\cite{Gu2023}. A key challenge in this context is twofold: firstly, trajectory data of sufficient quality is required to perform inference. Specifically, cells disappearing and reappearing from the tracking are a problem when inferring particle-based interactions, since all cells that are present should be considered to infer the interactions. Secondly, a formulation of Eq.~\ref{expansion_interacting} has to be found that is on the one hand flexible enough to capture potentially complex types of cell-cell interactions (which for instance do not have to obey radial symmetry), and on the other hand be restricted enough to allow accurate inference. Here, combining stochastic inference with sparsity constraints may be a way forward~\cite{Brunton2015}.

To test the potential symmetries of cell-cell interactions, a data-driven approach for cells in 2D monolayers using attention neural networks was recently proposed~\cite{LaChance2022} (Fig.~\ref{fig:collective}M,N). This approach detects how predictive the behaviour of neighbouring cells is for the behaviour of a given cell. In the case of radially symmetric interactions, such as in a Viscek-type alignment model, this attention map should be radially symmetric. However, this approach revealed that the single-cell response of fibroblast and epithelial cells are mainly affected by interactions with the neighbours ahead of them in the direction of motion (Fig.~\ref{fig:collective}N). In contrast, the interactions appeared isotropic in cancer cell collectives, showing how different cell types may exhibit different types of interacting symmetries.

Understanding cell-cell interactions becomes significantly more complicated in 3D systems, where cell migration often occurs on complex, curved surfaces. Therefore, data-driven theoretical approaches to these systems have primarily relied on a tissue-level or continuum description, rather than cell-resolved analysis as considered in the previous examples. Based on this, data-driven approaches for these systems often seek to decompose the dynamics into a low-dimensional set of modes, which we discuss next..

To study 3D collective migration \textit{in vitro}, minimal systems include cylindrical and spherical confinements (Fig.~\ref{fig:collective}G). In the case of spheroids, migrating cells usually setup global rotations of the tissue~\cite{Wang2013a,Chin2018,Palamidessi2019,Brandstatter2023,Tan2022}. Based on experimental cell trajectories, these rotational velocity fields could be decomposed into the basic mode of a rotational velocity field and the fluctuations in the co-rotating frame~\cite{Brandstatter2023}. In cancer organoids, this revealed travelling velocity waves with vortex flows~\cite{Brandstatter2023}, while in pancreas spheroids, a chiral velocity field was identified~\cite{Tan2022}. These collective modes could then be recapitulated with active cell migration models confined to the sphere, demonstrating that these modes are a generic response of active polar dynamics of cells to curvature.

The problem of inferring cell-cell interactions becomes more complex in \textit{in vivo} systems such as developing embryos. A popular model organism for cell migration in embryogenesis is zebrafish (Fig.~\ref{fig:collective}I). During zebrafish gastrulation, the tissue performs major rearrangements using a range of of biophysical processes, including guidance of cells by self-generated gradients~\cite{Stock2023}, motility-driven unjamming~\cite{Pinheiro2022}, and ECM-independent cell migration~\cite{Tavano2023}. While these various processes have been addressed with bottom-up active particle models, learning models from such embryo data could provide insight into the collective dynamics of this complex system. This was recently done at the scale of the entire embryo by decomposing the motion of the cells into a set of low-dimensional `developmental modes'~\cite{Romeo2021} (Fig.~\ref{fig:collective}O-Q). Specifically, density and flow fields were decomposed into a combination of basis functions (analogously to the basis functions for dynamical terms (Eq.~\ref{expansion})), using the spherical harmonics due to the shape of the system. The dynamics of these modes could then be described with  equations of motion identified by sparse regression. This allowed direct inference of a hydrodynamic model, revealing similarities between whole-embryo cell migration and active Brownian particle dynamics on curved surfaces.

In addition to cell migration, tissue flows due to active stresses exerted between cells are a key feature of morphogenesis, such as in \textit{Drosophila} gastrulation (Fig.~\ref{fig:collective}H). A common approach here is to develop continuum tissue mechanics models of embryos that match constitutive relations of active materials with experimental observations~\cite{Etournay2015,Morita2017,Streichan2018,Muenster2019}. Furthermore, recent work has proposed an approach to infer the active tensions driving tissue flow from the observed cell geometry based on the assumption of force balance~\cite{Noll2017,Brauns2023}. This allows a mode decomposition of the tissue dynamics into the dynamics of junctional angles (determined by tension) and cell areas (isogonal modes) (also refer to more general discussion in section~\ref{sec:learn_coll}). This also allowed disentangling active vs passive T1 transitions in the tissue, which have been shown to enable convergence extension-movement in large-scale tissue deformation processes such as during gastrulation~\cite{Ioratim-Uba2023,Sknepnek2023,Brauns2023}. A key challenge to test how well these models are constrained and their predictive power is whether one can for instance predict mutants, generalize the findings to other organisms, or make predictions for new experiments.

As a perspective for future research using data-driven approaches for interacting cells, we foresee two primary directions. On the one hand, data-driven approaches for cell-cell interactions could provide an avenue to better understand how molecular processes control interacting behaviours. One possible approach is to pharmacologically target molecular components that are known to be important in controlling cell-cell interactions, and then inferring the resulting change in dynamics. This may provide a way to identify the link between individual components with their role in the emergent behaviour. For example, a key question raised by the inference on two-cell collisions is how the separate positional and effective frictional components are controlled by molecular components. In particular, it is unclear what underlying mechanism controls the switch from friction or anti-friction interactions observed in non-cancerous and cancerous cells, respectively. Candidates are E-cadherin mediated cell-cell junctions, which are downregulated in cancer cells~\cite{Milano2016}, or ephrins, which play a key role in cell-cell recognition~\cite{Mayor2010}. Furthermore, to understand the emergence of the repulsive interaction between cells, which is responsible for Contact Inhibition of Locomotion, polarity cues, such as Rho GTPases, could be perturbed. These components are likely important in how cells change their direction of motion, an important process in the reversal events associated with CIL~\cite{Scarpa2013}. Thus, combining interaction inference with molecular perturbation in cell pair collision experiments could provide an avenue to link mechanisms and behaviour in interacting cellular systems. 

On the other hand, learning cellular interactions could help constrain active matter models which can then be used to make predictions for new experiments and biological systems beyond the dataset the model was trained on. For instance, inferring cell-cell interactions from pairs of colliding cells (Fig.~\ref{fig:collective}A) allows making predictions for these types of cells in more complex collective systems (Fig.~\ref{fig:collective}B-G). This would allow testing whether collective systems are explainable based on two-body interactions, reciprocal interactions, and whether these interactions exhibit adaptation to their environment.

Taken together, these examples and perspectives show how inference of cell-cell interactions can provide an important route towards understanding the active matter physics of interacting cells.

\section{Outlook}
\label{sec:outlook}

In this review, we have discussed how data-driven approaches make it possible to learn dynamical models of single and collectively migrating cells directly from experiments. The first part of this review on single-cell dynamics contained three main themes: how to infer and conceptualize models of cell behaviour in unstructured (free 2D) and structured (confining) systems; how to generalize these models to account for temporal and cell-to-cell variability in behaviour; and how to relate these behaviours to bottom-up models and underlying molecular mechanisms. While significant progress has been made in recent years to address these problems, much remains unknown. The basic molecular mechanisms that drive migration are increasingly well understood, but an integrated understanding of how these mechanisms interplay to set the emergent stochastic behaviours of cells at long time-scales still remains elusive. Furthermore, it remains unclear how the motility and cytoskeletal machinery of cells respond to external confinements at the molecular level, and how these responses determine the emergent behaviour.

So far, data-driven models of cell migration dynamics, from persistent random motion, to confined cell migration and interacting cells, have often been limited to the treatment of low-dimensional sets of cellular degrees of freedom, such as cell nucleus trajectories. These models could in the future provide important constraints for bottom-up biophysical models (see section~\ref{sec:bottomup_models}). This could yield more interpretable, yet strongly data-constrained descriptions, of the underlying mechanisms. Another exciting avenue is to take data-driven approaches to a more mechanistic level by tracking and analyzing subcellular degrees of freedom, such as actin flows, polarity markers, or traction forces (see section~\ref{sec:beyond_traj}).

In the second part of the review, we discussed how data-driven approaches for cell-cell interactions could provide an avenue to better understand how molecular processes control interacting behaviours by performing model inference on a range of genetic and pharmacological perturbations of underlying migration and interaction mechanisms. Inferring these interactions for a range of cell types and extra-cellular environment could reveal new and unexpected types of interactions, such as non-reciprocal, adaptive, or time-dependent terms. The inferred interactions could then allow simulation and prediction of the collective dynamics at larger scales. This would allow testing whether two-body dynamics are predictive of many-body dynamics. In systems where the inferred two-body interactions are non-standard, this would also allow exploring their consequences in more complex systems. Moreover, applying inference directly to collective migration datasets, allowing to disentangle the separate contributions of single-cell motility, cell-cell interactions, cell proliferation and external confinements.

An attractive perspective is that beyond providing tools for building predictive physical models and constraining underlying mechanisms, data-driven approaches can also help provide new conceptual insights. Cell migration requires the coordination of a multitude of molecular players on the cell level and of the behaviours of a large number of cell on the tissue scale, such as the equilibration to force balance in epithelia. Achieving such coordination has been proposed to occur on low-dimensional `slow' manifolds in high-dimensional systems in a variety of contexts. An example for this are cell fate decisions, where high-dimensional expression profiles can be described by effective two-dimensional dynamical systems~\cite{Rand2021,Saez2021}, as originally introduced through the idea of Waddington's landscape~\cite{Waddington1957}. Similar principles could potentially play a role in how cells and tissues coordinate their behaviours, and data-driven inference could allow us to identify the low-dimensional dynamical systems and attractor manifolds governing these behaviours.

Taken together, these perspectives demonstrate how data-driven approaches have the potential to address key open questions in single and collective cell migration. A common thread in these ideas is that by applying data-driven inference to experimental datasets, we can complement models developed from the bottom up, by inferring models directly from data. Based on these inferred dynamics, we can then attempt to constrain underlying mechanisms, and predict emergent behaviours of the system.

\section*{Acknowledgements}
This work was supported by the Deutsche Forschungsgemeinschaft (DFG, German Research Foundation) - Project-ID 201269156 - SFB 1032 (Project B12). D.B.B. was supported by a NOMIS Fellowship and an EMBO Fellowship (ALTF 343-2022). We thank Joachim R\"adler, Alexandra Fink, Erwin Frey, Pierre Ronceray, Ricard Alert, Edouard Hannezo, Henrik Flyvbjerg, Ulrich Schwarz, Joshua Shaevitz, Greg Stephens, Andrea Cavagna, Grzegorz Gradziuk, Fridtjof Brauns, Nikolas Claussen, Tom Brandst\"atter, Johannes Flommersfeld, Christoph Schreiber, Nicolas Arlt, Matthew Schmitt,  Joris Messelink, Federico Gnesotto, Federica Mura, Bram Hoogland, Manon Wigbers, Isabella Graf, Jessica Lober, and many others for inspiring discussions. We also thank Claudia Flandoli for the artwork in Figs.~\ref{fig:overview}, \ref{fig:confined}, \ref{fig:models} and~\ref{fig:collective}.


\bibliographystyle{ieeetr}
\bibliography{/Users/D.Brueckner/Zotero/my_library}

\end{document}